\documentclass[12pt]{article}

\usepackage{epsfig,amsmath,amssymb,verbatim,mathrsfs}

\def\beq{\begin{eqnarray}}
\def\eeq{\end{eqnarray}}
\def\bea{\begin{eqnarray}}
\def\eea{\end{eqnarray}}

\def\tev{\, {\rm TeV}}
\def\gev{\, {\rm GeV}}

\newcommand{\gsim}{\lower.7ex\hbox{$\;\stackrel{\textstyle>}{\sim}\;$}}
\newcommand{\lsim}{\lower.7ex\hbox{$\;\stackrel{\textstyle<}{\sim}\;$}}

\def\stilde{\widetilde}

\newcommand{\newc}{\newcommand}
\newc{\Nc}{N_{c}}
\newc{\CG}{C_G}
\newc{\gp}{g'}
\newc{\stopi}{\stilde t_i}
\newc{\sboti}{\stilde b_i}
\newc{\staui}{\stilde \tau_i}
\newc{\stopj}{\stilde t_j}
\newc{\sbotj}{\stilde b_j}
\newc{\stauj}{\stilde \tau_j}
\newc{\stopI}{\stilde t_1}
\newc{\stopII}{\stilde t_2}
\newc{\sbotI}{\stilde b_1}
\newc{\sbotII}{\stilde b_2}
\newc{\stauI}{\stilde \tau_1}
\newc{\stauII}{\stilde \tau_2}
\newc{\sstop}{s_{t}}
\newc{\cstop}{c_{t}}
\newc{\ssbot}{s_{b}}
\newc{\csbot}{c_{b}}
\newc{\sstau}{s_{\tau}}
\newc{\cstau}{c_{\tau}}
\newc{\Sstop}{s_{2t}}
\newc{\Cstop}{c_{2t}}
\newc{\Ssbot}{s_{2b}}
\newc{\Csbot}{c_{2b}}
\newc{\Sstau}{s_{2\tau}}
\newc{\Cstau}{c_{2\tau}}
\newc{\salpha}{s_\alpha}
\newc{\calpha}{c_\alpha}
\newc{\Calpha}{c_{2\alpha}}
\newc{\Salpha}{s_{2\alpha}}
\newc{\sbetapm}{s_{\beta_\pm}}
\newc{\cbetapm}{c_{\beta_\pm}}
\newc{\Sbetapm}{s_{2 \beta_\pm}}
\newc{\Cbetapm}{c_{2 \beta_\pm}}
\newc{\sbetaO}{s_{\beta_0}}
\newc{\cbetaO}{c_{\beta_0}}
\newc{\SbetaO}{s_{2 \beta_0}}
\newc{\CbetaO}{c_{2 \beta_0}}
\newc{\vu}{v_u}
\newc{\vd}{v_d}
\newc{\seL}{\stilde e_L}
\newc{\smuL}{\stilde \mu_L}
\newc{\seR}{\stilde e_R}
\newc{\smuR}{\stilde \mu_R}
\newc{\suL}{\stilde u_L}
\newc{\sdL}{\stilde d_L}
\newc{\suR}{\stilde u_R}
\newc{\sdR}{\stilde d_R}
\newc{\scL}{\stilde c_L}
\newc{\ssL}{\stilde s_L}
\newc{\scR}{\stilde c_R}
\newc{\ssR}{\stilde s_R}
\newc{\snue}{\stilde \nu_e}
\newc{\snumu}{\stilde \nu_\mu}
\newc{\snutau}{\stilde \nu_\tau}
\newc{\Gpm}{G^\pm}
\newc{\Hpm}{H^\pm}
\newc{\FFbS}{\overline{FF}S}
\newc{\FFbV}{\overline{FF}V}
\newc{\FSS}{F_{SS}}
\newc{\FSSS}{F_{SSS}}
\newc{\FFFS}{F_{FFS}}
\newc{\FFFbS}{F_{\overline{FF}S}}
\newc{\FSSV}{F_{SSV}}
\newc{\FVS}{F_{VS}}
\newc{\FVVS}{F_{VVS}}
\newc{\FFFV}{F_{FFV}}
\newc{\FFFbV}{F_{\overline{FF}V}}
\newc{\Fgauge}{F_{\rm gauge}}
\newc{\DRbarprime}{$\overline{\rm DR}'$ }
\newc{\DRbar}{$\overline{\rm DR}$ }
\newc{\MSbar}{$\overline{\rm MS}$ }
\newc{\Yu}{{\bf Y}_u}
\newc{\Yd}{{\bf Y}_d}
\newc{\Ye}{{\bf Y}_e}
\newc{\Au}{{\bf a}_u}
\newc{\Ad}{{\bf a}_d}
\newc{\Ae}{{\bf a}_e}
\newc{\bm}{{\bf m}}
\newc{\zhol}{Z^{\rm hol}}

\newc{\rwino}{r_{\tilde W}}
\newc{\rmu}{r_{\tilde H}}
\newc{\ra}{r_A}
\newc{\ccdot}{\!\cdot\!}

\newcommand{\nnmb}{\nonumber}

\newcommand{\met}{\,/\hspace{-0.25cm}E_T}

\begin{document}

\setlength{\baselineskip}{0.2in}



\begin{titlepage}
\noindent
\begin{flushright}
MCTP-06-32  \\
\end{flushright}
\vspace{1cm}

\begin{center}
  \begin{Large}
    \begin{bf}
Higgs Boson Exempt No-Scale Supersymmetry\\ 
and its Collider and Cosmology Implications \\
     \end{bf}
  \end{Large}
\end{center}
\vspace{0.2cm}
\begin{center}
\begin{large}
Jason L. Evans, David E. Morrissey,  James D. Wells \\
\end{large}
  \vspace{0.3cm}
  \begin{it}
Michigan Center for Theoretical Physics (MCTP) \\
Physics Department, University of Michigan, Ann Arbor, MI 48109

\vspace{0.1cm}
\end{it}

\end{center}

\center{\today}

\begin{abstract}

 One of the most straightforward ways 
to address the flavor problem of low-energy supersymmetry
is to arrange for the scalar soft terms to vanish simultaneously
at a scale $M_{c}$ much larger than the electroweak scale. 
This occurs naturally in a number of scenarios, such as  
no-scale models, gaugino mediation, and several models with
strong conformal dynamics. Unfortunately, the most basic version of
this approach that incorporates gaugino mass unification and zero scalar
masses at the grand unification scale is not compatible with collider and
dark matter constraints. However, experimental constraints
can be satisfied
if we exempt the Higgs bosons from flowing to zero mass value at
the high scale. We survey the theoretical constructions that allow this, 
and investigate the collider and dark matter consequences.  
A generic feature is that the sleptons are relatively light.  
Because of this, these models frequently give a significant contribution
to the anomalous magnetic moment of the muon, and neutralino-slepton 
coannihilation can play an important role in obtaining an acceptable 
dark matter relic density.  Furthermore, the light sleptons give rise 
to a large multiplicity of lepton events at colliders, including a 
potentially suggestive clean trilepton signal at the Tevatron,
and a substantial four lepton signature at the LHC.

\end{abstract}

\vspace{1cm}

\end{titlepage}

\setcounter{page}{2}

\tableofcontents



\section{Introduction}

 Supersymmetry is a well-motivated way to extend the standard model~(SM).
Most impressively, supersymmetry can stabilize the large disparity between
the size of the electroweak scale and the Planck scale~\cite{Martin:1997ns}.  
In addition, the minimal supersymmetric extension of the 
SM~\cite{Dimopoulos:1981zb}, the MSSM, leads to 
an excellent unification of the $SU(3)_c$, $SU(2)_L$, and $U(1)_Y$
gauge couplings~\cite{Dimopoulos:1981yj} near $M_{GUT}=2\times 10^{16}$~GeV, 
a scale that is large enough that grand-unified theory~(GUT) induced 
nucleon decay is not a fatal problem.  The MSSM also contains a new 
stable particle if $R$-parity is an exact symmetry.  This new stable 
particle can potentially make up the dark matter.

  The main obstacles facing supersymmetric extensions of the SM  
come from the requirement that supersymmetry be (softly) broken.
To preserve the natural supersymmetric hierarchy between $M_W$ 
and $M_{\rm Pl}$, every MSSM operator that breaks supersymmetry should
be accompanied by a dimensionful coupling of size less than
about a TeV.  However, for a generic set of soft terms of this size, 
consistent with all the symmetries of the theory, the amount of flavor 
mixing and CP violation predicted by the model is much greater than 
has been observed.  Instead, the experimental constraints require that 
the soft terms be nearly flavor-diagonal in the super-CKM 
basis~\cite{Gabbiani:1996hi}, and that nearly all the independent 
CP violating phases be very small~\cite{Gabbiani:1996hi,Dugan:1984qf}, 
or finely-tuned to cancel~\cite{Brhlik:1998zn}.  From the low-energy 
perspective, it is not clear why this should be so.  

  A number of approaches to the supersymmetric flavor and CP problems have
been put forward, such as adding new flavor symmetries to the MSSM~\cite{flav},
or mediating supersymmetry breaking through gauge 
interactions~\cite{gauge-med} or the superconformal anomaly~\cite{ano-med}.
These models also face new difficulties.
New flavor symmetries typically require additional matter fields
and hence the complications that go with them.  Gauge mediation generates 
flavor-universal soft masses and trivially small $A$ terms, 
but does not fully solve the CP problem, and makes it difficult 
to generate both the $\mu$ and $B\mu$ terms with the correct size.  
Anomaly mediation in its most simple form suffers from tachyonic slepton 
soft masses~\cite{ano-med}.
A more radical approach to the flavor and CP problems is to 
push the scale of the soft supersymmetry breaking scalar couplings to
be much larger than the electroweak scale~\cite{Wells:2003tf,splitsusy}. 
If this is done while keeping the gauginos relatively light,
it is possible to preserve gauge unification and a good dark matter 
candidate~\cite{moresplit}.
Of course, supersymmetry would no longer directly solve the gauge hierarchy
problem if the scalar superpartners are very heavy. 

  Another way to address the supersymmetric flavor problem, and the one 
we consider in the present work, is to have the soft scalar 
masses and the $A$ terms vanish simultaneously at a scale 
$M_{c}$~\cite{Dine:1990jd}.  If this scale is much larger than 
the electroweak scale, and if the gaugino masses do not vanish at 
$M_{c}$, non-zero values for the scalar soft terms will be generated 
by radiative effects as the theory is evolved to lower energies.  
Since the scalar soft terms thus induced are family-universal, 
the resulting soft spectrum does not have a flavor problem.  
The supersymmetric CP problem is also improved but not solved 
by this approach.  
Besides the CKM phases, the only remaining phases are those of the 
gaugino soft masses, and the $\mu$ and $B\mu$ terms.  If the gaugino
soft mass phases are universal, there are three new phases of which
two can be removed by making field redefinitions~\cite{Dugan:1984qf}.
The remaining phase can be eliminated as well within 
particular models~\cite{gaugino-med}.

  Near-vanishing soft scalar terms can arise in a number of ways.
The canonical examples are the no-scale models of gravity mediated
supersymmetry breaking.  In these models, the absence of scalar
soft terms is related to the flatness of the hidden sector potential
that allows the gravitino mass to be determined by loop-corrections
due to light fields~\cite{noscale}.
A more recent construction that leads to near-vanishing
soft scalar operators is gaugino-mediated supersymmetry
breaking~\cite{gaugino-med}.  Here, the MSSM chiral multiplets 
are separated from the source of supersymmetry breaking by an 
extra-dimensional bulk, while the gauge multiplets propagate in the bulk.  
Locality in the extra dimension(s) leads to gaugino masses that 
are much larger than the scalar soft terms.  Small scalar soft terms can also
be obtained from strong conformal dynamics in either the visible
or the hidden sectors.  In these constructions, the conformal running 
suppresses the scalar soft terms exponentially 
relative to the gaugino soft masses~\cite{confseq}.

  The main difficulty with very small input scalar soft masses
is that the lightest SM superpartner particle is usually a 
mostly right-handed slepton, which can be problematic for cosmology.  
This is nearly always the case if gaugino universality is assumed 
to hold above $M_{GUT}$, and $M_{c}\leq M_{GUT}$.  
On the other hand, if $M_{c}$ is an order of magnitude or more
above $M_{GUT}$ (with gaugino universality), the lightest superpartner 
becomes a mostly Bino neutralino.  A viable low-energy spectrum can
be obtained in this way~\cite{Schmaltz:2000gy,Schmaltz:2000ei}.
For $M_{c} \leq M_{GUT}$, a neutralino LSP can be obtained
by relaxing the requirement that all soft scalar terms vanish
at $M_{c}$.  One such generalization that does not re-introduce 
a flavor problem is to allow the Higgs soft masses $m_{H_u}^2$ 
and $m_{H_d}^2$ to be non-zero at $M_{c}$.  These soft masses
contribute to the running of the slepton masses through a hypercharge
Fayet-Iliopoulos $D$-term, and can push the slepton masses above 
that of the lightest neutralino~\cite{Schmaltz:2000ei,Kaplan:2000av,
Buchmuller:2005ma}.

  In the present work, we study the phenomenology
of the MSSM subject to vanishing scalar soft terms.  We generalize
our study by including non-vanishing Higgs boson soft masses, as 
well as a Higgs boson bilinear $B$ term.  This does not reintroduce
a flavor problem.  Inspired by grand-unified
theories, we take our input scale to be $M_{GUT}$, and demand
universal gaugino masses at this scale.\footnote{For the case
of non-universal gaugino masses but vanishing Higgs soft terms, 
see~\cite{Komine:2000tj}.}  After imposing consistent electroweak 
symmetry breaking, the independent parameters of this theory,
which we call Higgs Exempt No-Scale (HENS) supersymmetry, are 
\beq
M_{1/2},~~\tan\beta,~~m_{H_u}^2,~~m_{H_d}^2,~~sgn(\mu),
\eeq  
where $M_{1/2}$ is the universal gaugino mass at $M_{GUT}$,
$\tan\beta$ is the ratio of the Higgs boson expectation values,
$m_{H_u}^2$ and $m_{H_d}^2$ are the soft Higgs masses at $M_{GUT}$,
and $sgn(\mu)$ is the sign of the supersymmetric Higgs bilinear $\mu$ term.

  The outline of this paper is as follows.
In Section~\ref{theo} we motivate this scenario, and show
how it can emerge in a number of different ways.  Section~\ref{mass} 
discusses the low-energy mass spectrum of the model, the acceptable regions of
parameter space, and the most important constraints on these regions.
In Section~\ref{dark} we investigate the prospects for dark matter
in the model, and discuss some of the potential signatures this
might induce.  Section~\ref{coll} contains an investigation of some
of the potential collider signatures of the model at the Tevatron
and the LHC, as well as a discussion of the discovery prospects at
these machines.  Finally, Section~\ref{conc} is reserved for our conclusions.

  While this work was in preparation, we became aware of Refs.~\cite{
Buchmuller:2005ma,Buchmuller:2006nx},
which investigate the mass spectrum and dark matter prospects of
gaugino mediation with unsuppressed Higgs soft masses.  There is some
overlap between their work and the material in Sections~\ref{mass} and
\ref{dark} of this paper.  However, unlike Refs~\cite{Buchmuller:2005ma,
Buchmuller:2006nx}, we allow negative Higgs soft squared masses at 
the input scale~\cite{Feng:2005ba}, 
and have a more extensive discussion of the 
phenomenological constraints.  On the other hand, they consider 
specific details of gaugino mediated models, and discuss the possibility 
of gravitino dark matter in more detail than we do.

\section{Generating Small Scalar Soft Terms}
\label{theo}
  
  Our primary motivation for considering models with small
scalar soft terms is data driven: they provide a simple and
elegant reason for small flavor violations induced by supersymmetry.
Even so, it is comforting that this framework can arise from a
number of theoretical constructions.  In this section we describe
some of these models, and discuss how they can be modified
to allow for non-vanishing soft masses for the Higgs fields.

\subsection{Models} 

  Vanishing scalar soft terms have traditionally been
associated with no-scale models~\cite{noscale}.  These models are
attractive because the gravitino mass, and therefore the scale
of supersymmetry breaking, is determined dynamically.
The basic assumption underlying no-scale constructions 
is that the effective superspace K\"ahler density
and superpotential have the form~\cite{Cremmer:1983bf,Ellis:1983sf,
Ellis:1983ei,Ellis:1984bm}
\bea
\mathcal{F} &=& -3\,M_{\rm Pl}^2 + f(X) + f^{\dagger}(X^{\dagger}) 
+ g(\Phi,\Phi^{\dagger}),\label{g-noscale}\\
\mathcal{W} &=& W(\Phi),\nnmb
\eea
where $X$ is a hidden sector field, $\Phi$ represents a visible
sector field, and $W$ is a holomorphic cubic function.  
With this form of the K\"ahler density and superpotential, 
the tree-level potential along the direction of the hidden sector 
field $X$ is flat, the gravitino mass, $m_{3/2}$ is undetermined,
and no soft terms are generated for the visible sector scalars.
Supersymmetry breaking is communicated to the visible sector
by non-trivial $X$-dependent gauge kinetic functions, which generate
gaugino soft masses on the order of $m_{3/2}$.  The one-loop corrections
from the gauginos lift the potential in the $X$ direction and 
fix the value of $m_{3/2}$ to lie close to the electroweak scale,
which is determined dynamically
by the large top Yukawa coupling~\cite{Ellis:1983sf}.

  It is difficult to maintain $m_{3/2} \sim M_W$ in no-scale models
if there are other larger scales in the theory because of the radiative
corrections these contribute to the effective potential for 
$m_{3/2}$~\cite{Ellis:1984bm}.  
If such large scales exist, such as in a GUT, the heavy sector must 
be completely sequestered from the supersymmetry breaking, since in 
the supersymmetric limit, they do not alter the effective potential.
Within a GUT where the Higgs superfields are components of 
complete multiplets that also contain heavy fields,
it is therefore essential to prevent these GUT multiplets 
from obtaining a supersymmetry breaking mass.
Having separated the Higgs in this way, it is natural to sequester
the other chiral multiplets as well.  This is the origin of the vanishing
scalar soft terms in no-scale models.  Gaugino masses can be induced by
a non-minimal kinetic function for those and only those components of the GUT
vector multiplet that remain light.  
Thus if the Higgs multiplets are components 
of a larger GUT multiplet, of which some components develop 
GUT scale masses, it is not possible to generate soft masses for 
the Higgs fields without destabilizing $m_{3/2}$.  On the other hand,
soft Higgs masses might be possible in more general unification scenarios in 
which the Higgs fields do not belong to complete 
GUT multiplets~\cite{Weiner:2001pv}.

  The form of the no-scale K\"ahler density and superpotential, 
Eq.~(\ref{g-noscale}), is an input to these models.  Such a form does 
arise to lowest order in several string- and M-theory 
constructions, but is typically corrected 
at higher orders~\cite{noscalestring}.
More generally, a superspace K\"ahler density in which the visible and
hidden sectors appear as disjoint terms,
as in Eq.~(\ref{g-noscale}), is said to be \emph{sequestered}~\cite{ano-med}.  
A sequestered K\"ahler density and superpotential
guarantees that no direct soft terms are generated.  This is a
necessary ingredient for anomaly mediation~\cite{ano-med}. 
 
  Complete sequestering can be obtained geometrically by confining
the visible and hidden sectors to branes separated by an 
extra-dimensional bulk~\cite{ano-med}.  A partial sequestration can also be 
realized if the MSSM gauge multiplets are allowed to propagate in 
the bulk, as in {gaugino mediated supersymmetry breaking}~\cite{gaugino-med}.  
If so, the gauginos will develop soft masses through their local 
couplings to the hidden sector, while the soft terms of the chiral 
multiplets will only be generated by loops passing across the bulk.  
The resulting supersymmetric spectrum at the compactification scale 
of the extra dimension is therefore close to the one we are interested in:
non-zero gaugino masses, and much smaller soft scalar terms. 
By allowing the Higgs multiplets to propagate in the bulk,
non-zero Higgs soft terms can be generated as well.  
This is perhaps the simplest way to realize the HENS models 
we shall consider.

  Sequestering can also be realized in four dimensions through 
strongly-coupled conformal dynamics in the hidden sector~\cite{confseq}.  
Even without strong conformal running, the contributions to the gaugino 
masses and the trilinear $A$ are naturally suppressed relative to 
the gravitino mass if there are no singlets in the hidden sector.  
This is not true for the scalar soft masses, which are generated 
by terms like
\beq
\mathscr{L} \supset \int d^4\theta\;\frac{c_{ij}}{M_*}\Phi^{\dagger}_i\Phi_j\,
X^{\dagger}X \to \frac{c_{ij}|F_X|^2}{M_*}\phi_i^*\phi_j,
\label{softmass}
\eeq
where $M_*$ is the messenger scale, $X$ is a hidden sector field,
and $\Phi_i$ is a visible field.  Operators of the form of Eq.~(\ref{softmass})
can be suppressed relative to the gravitino mass by strong conformal 
dynamics in the hidden sector that couples at the renormalizable level 
to $X$~\cite{Dine:2004dv}.  
If all such soft mass contributions are sufficiently suppressed,\footnote{
All the soft masses will be suppressed provided there are no unbroken,
non-anomalous global symmetries in the hidden sector.  The amount
of suppression depends on the range over which the conformal running
takes place and the beta-function of the corresponding gauge coupling.}
and if there are no hidden sector singlets, the visible sector is 
sequestered and the leading contribution to the soft terms comes 
from anomaly mediation.  By allowing singlets in the hidden sector,
both gaugino masses and $A$ terms can be generated that are much
larger than the anomaly-mediated terms~\cite{Dine:2004dv}.  This is 
close to, but not quite, the spectrum we are interested in, because
it is unclear how to avoid suppressing the Higgs soft masses 
while squashing the rest.

  A partial sequestration of soft terms, as well as an explanation
for the Yukawa hierarchy, can also be obtained from strong conformal 
dynamics in the visible sector~\cite{Nelson:2000sn,Nelson:2001mq,
Kobayashi:2001kz}.  In these constructions, there is a new gauge group 
$G_c$ that approaches a strongly-coupled fixed point in the IR.  
The MSSM fields are not charged under $G_c$, but they do couple to 
fields that are through cubic operators in the superpotential.  
As the theory flows towards the fixed point, the MSSM
fields develop large anomalous dimensions which suppress their 
corresponding (physical) Yukawa couplings.  
Since different (linear combinations of) fields develop distinct 
anomalous dimensions, related to their effective superconformal 
$R$ charges, a Yukawa hierarchy can be generated in this 
way~\cite{Nelson:2000sn}.  
The conformal running also produces a general suppression of the soft 
scalar masses, as well as a hierarchy of trilinear $A$ terms that 
mirrors the Yukawa couplings~\cite{Nelson:2000sn,Nelson:2001mq,
Kobayashi:2001kz,Kobayashi:2002iz}.
Conversely, the gaugino masses are largely unaffected because they 
do not couple directly to the strongly-coupled sector.  The third
generation multiplets and the Higgs multiplets
must also be shielded from the conformal running effects to avoid 
suppressing the top quark Yukawa coupling.  As a result, the third 
generation and the Higgs soft masses do not get suppressed.
Thus, the spectrum from visible sector conformal running is similar
to one we shall consider, but augmented by third generation soft masses
and $A$ terms.\footnote{There may also be additional contributions to the
soft masses if there are non-anomalous, continuous, abelian global symmetries.}
We expect the phenomenology of both scenarios to be similar over much
of the allowed parameter space.

\subsection{The Scale of $M_c$}

  From the discussion above, we see that a HENS soft mass spectrum
can arise from gaugino mediation with the Higgs 
multiplets in the bulk, or from conformal running in the hidden sector
up to additional contributions to the third generation states.
Before proceeding, however, let us comment on our choice of  
$M_{GUT}$ as the input scale for the soft spectrum.   
In gaugino mediation, the input scale is on the order 
of the compactification scale, $M_c$.\footnote{
$M_c := 1/R$, is less than the cutoff of the theory.}
For visible-sector conformal running, the input scale for the 
soft spectrum is the scale at which the conformal running ceases,
which we will also call $M_c$.  Our motivation to set $M_c = M_{GUT}$
is partly conventional, but is also motivated by gauge unification
and our wish to strongly suppress the scalar soft masses. 

  In both cases, gauge unification can be preserved with 
$M_c < M_{GUT}\sim 2\times 10^{16}~\gev$, but the process will be 
more complicated than in the standard picture.  In gaugino mediation, 
Kaluza-Klein states appear above $M_c$ and can lead to an accelerated 
power-law running~\cite{Dienes:1998vh}.  
With conformal dynamics in the visible sector, the SM gauge coupling 
beta functions will be modified by the large anomalous dimensions 
of the MSSM fields.  Gauge unification will still occur, albeit at 
a lower scale, provided the conformal dynamics respects a global 
symmetry into which the SM gauge group can be embedded~\cite{Nelson:1996km}.
Thus, in each case having $M_c$ below $M_{GUT}$ can induce an effective 
unification of the SM gauge couplings below the apparent unification 
scale $M_{GUT}\simeq 2\times 10^{16}~\gev$.  This is problematic for many
GUT completions of the MSSM, which predict baryon and lepton number
violation.  Typically, some additional structure is needed if $M_c$
is much smaller than $M_{GUT}$.  This motivates us to consider
$M_c\geq M_{GUT}$.

  It is clear that gauge unification can also be maintained with
$M_c \geq M_{GUT}$.  If $M_c$ is larger than $M_{GUT}$, 
the renormalization group running from $M_c$ down to $M_{GUT}$ will induce 
non-vanishing (flavor-universal) soft masses at $M_{GUT}$.
The size of these corrections from running above $M_{GUT}$
depends on the precise $GUT$ completion of the theory,
but even for minimal $GUT$ models they can be significant,
on the order of~\cite{Schmaltz:2000gy,Schmaltz:2000ei}
\bea
\Delta A &\simeq& \frac{2\alpha_{G}}{\pi}\,C_A\;
\ln\left(\frac{M_c}{M_{GUT}}\right)\,M_{1/2},\label{abovemg}\\
\Delta m^2 &\simeq& \frac{2\alpha_{G}}{\pi}\,C_{m^2}
\ln\left(\frac{M_c}{M_{GUT}}\right)\,M_{1/2}^2,\nnmb
\eea
where $\alpha_G \simeq 1/24$ is the GUT coupling,
and $C_A$ and $C_{m^2}$ are dimensionless constants 
on the order of or slightly larger than unity.
These contributions can be large enough for an acceptable
low-energy spectrum to be obtained~\cite{Schmaltz:2000gy,Schmaltz:2000ei}.

  On the other hand, in both gaugino mediation and conformal sequestering,
$M_c$ cannot be more than about an order of magnitude above $M_{GUT}$
because the suppression of soft terms (and Yukawa 
couplings) requires a separation of scales.  Let $M_{in}$ be the scale
at which conformal running begins in the case of conformal dynamics,
or the $UV$ cutoff of the extra-dimensional gauge theory in gaugino mediation.
Presumably $M_{in} \leq M_{\rm Pl} = 2.4\times 10^{18}~\gev$.  
The amount of suppression of the soft scalar terms from conformal 
running is expected to be an order-one power of $M_c/M_{in}$, 
whereas the required suppression is typically on the order 
of $10^{-4}$~\cite{Nelson:2001mq}.  An even stronger upper bound
on $M_{c}$ can be obtained if the conformal dynamics are responsible
for the small electron Yukawa coupling as in Ref.~\cite{Nelson:2000sn}.
The condition for this is
\beq
y_e \simeq \left(\frac{3\times 10^{-6}}{\cos\beta}\right) \simeq 
\left(\frac{M_c}{M_{in}}\right)^{(\gamma_{L}+\gamma_{E})/2},
\eeq
where $\gamma_i$ denote the anomalous dimensions of $L$ and $E^c$,
which are generally smaller than 2.  If we take this bound seriously,
$M_c$ can be at most only slightly larger than $M_{GUT}$.
In gaugino mediation, flavor-mixing contact interactions between 
the MSSM chiral multiplets and hidden sector operators, arising from 
bulk states with masses above the UV cutoff scale, are suppressed by a factor
of $\exp(-M_{in}/M_c)$~\cite{gaugino-med}.  Again this factor must
be less than about $10^{-4}$ to avoid various experimental 
flavor constraints, which translates into $M_{c}$ being within an 
order of magnitude larger than $M_{GUT}$ (for $M_{in} = M_{\rm Pl}$).

Given the above considerations,  we will set $M_c = M_{GUT}$ throughout 
this paper, and not concern 
ourselves with the precise mechanism by which the scalar soft terms 
are suppressed.  While beyond the scope of the present work, it also 
interesting to speculate that the breaking of the GUT symmetry is 
related to the geometry of the extra dimension, or the escape 
from conformal running.  Such a construction would further 
justify our choice of $M_c = M_{GUT}$.  Finally, let us also note 
that within particular models there is typically some residual 
flavor violation due to an incomplete suppression of the scalar terms
at $M_c$.  The amount of flavor suppression can be close to the
level probed by current experiments.  However, without specifying
a particular model, it is not possible to perform an analysis
of the constraints due to flavor physics. Thus, we assume as our starting
point that at the scale $M_c=M_{GUT}$, all scalar masses except those
of the Higgs bosons are precisely zero and that corrections to that assumption
are inconsequential to the phenomenology discussed below.

\section{Mass Spectrum and Constraints}
\label{mass}

\subsection{One-Loop Analysis}

  The essential features of the HENS mass spectrum are well illustrated
by a simple one-loop analysis.  At this order, the ratio $M_a/g_a^2$, 
$a=1,2,3$, is scale invariant for all three gaugino masses.  
If the gauge couplings unify and the gaugino masses are universal 
at $M_{GUT}$, it follows that at lower scales Q, $M_a(Q) = [g_a(Q)/g_{GUT}]^2
M_{1/2}$.  For $Q = 1\,\tev$, this gives
\beq
M_1 \simeq (0.43)\,M_{1/2},~~~M_2 \simeq (0.83)\,M_{1/2},~~~M_3 
\simeq (2.6)\,M_{1/2}.\label{2:m12}
\eeq

  The one-loop running of the scalar soft masses is 
given by~\cite{Martin:1997ns,Martin:1993zk}
\beq
(4\pi)^2\,\frac{dm_i^2}{dt} \simeq X_i -8\,\sum_a\,C^a_i\,g_a^2\,|M_a|^2
+ \frac{6}{5}\,g_1^2\,Y_i\,S,
\label{2:msoftrun}
\eeq
where $X_i$ depends on the soft masses and $A$ terms and is usually 
proportional to Yukawa couplings, $C_i^a$ is the quadratic Casimir
for the representation $i$ under gauge group $a$, and 
\beq
S = (m_{H_u}^2-m_{H_d}^2) + tr_F(m_Q^2-2m_U^2+m_E^2+m_D^2-m_L^2),
\label{2:sterm}
\eeq
with the trace above running over flavors.  

  At one-loop order, the RG equation for the $S$ term is particularly simple,
\beq
(4\pi)^2\,\frac{d S}{dt} = \frac{66}{5}g_1^2\,S.
\eeq
Because of this simple form, the effect of the $S$-term on the 
low-scale soft masses is to simply shift the value they would have with $S=0$
by the amount
\beq
\Delta m_i^2 = -\frac{Y_i}{11}\left[1-\left(\frac{g_1}{g_{GUT}}\right)^2\right]
\,S_{GUT} \simeq -(0.052)\,Y_i\,S_{GUT},
\eeq
where $S_{GUT} = (m_{H_u}^2-m_{H_d}^2)$ evaluated at $M_{GUT}$.

  Neglecting Yukawa effects, the low-scale slepton soft masses 
at $Q=1\,\tev$ are
\bea
m_L^2 &\simeq& \left[(0.68)\,M_{1/2}\right]^2 + \frac{1}{2}(0.052)\,S_{GUT},
\label{2:mslepton}\\ 
m_E^2 &\simeq& \left[(0.39)\,M_{1/2}\right]^2 - (0.052)\,S_{GUT}.
\eea
If mixing effects are small, the physical slepton masses will be
close to $\sqrt{m_L^2}$ and $\sqrt{m_E^2}$, up to the $U(1)_Y$ $D$-term 
contributions.  The mass of the lightest neutralino is usually
close to $M_1$ (under the assumption of gaugino universality)
unless $\mu$ is relatively small.

  Comparing Eq.~\eqref{2:mslepton} with Eq.~\eqref{2:m12}, we see that 
for $S_{GUT} \geq 0$, $m_E^2$ is less than $M_1$ and the lightest superpartner 
tends to be a mostly right-handed slepton.  On the other hand, 
if $S_{GUT} < 0$, the right-handed slepton soft mass is pushed 
up relative to $M_1$, allowing for a mostly Bino neutralino LSP.  
For $S_{GUT}$ very large and negative, the LSP can be a mostly 
left-handed slepton. Relative to the sleptons and the electroweak gauginos,
the squarks and gluino are very heavy because the $SU(3)_c$ gauge
coupling grows large in the infrared.

\subsection{Parameter Space Scans}

  To confirm the simple analysis given above, we have performed
a scan over the HENS parameter space using SuSpect~2.34~\cite{Djouadi:2002ze}.
This code performs the renormalization group running at two-loop order 
with one-loop threshold effects, and includes radiative and mixing corrections 
to the physical particle masses.  We take $\alpha_s(M_Z) 
= 0.118$~\cite{Yao:2006px} and $m_t = 171.4$~GeV~\cite{Brubaker:2006xn}
in our analysis.  For each model parameter point 
we require consistent electroweak symmetry breaking, and superpartner 
masses above the LEP~II and Tevatron bounds 
($m_{\chi_1^0},~m_{\tilde{\nu}} > 46$~GeV, $m_{\tilde{l}} > 90~\gev$, 
$m_{\chi_1^{\pm}} > 104~\gev$).
We also impose the lower-energy constraints
\bea
\Delta\rho &\in& [-8,\,24]\times 10^{-4}~~~~\cite{Yao:2006px}\\
BR(b\to s\gamma) &\in& 
[3.0,\,4.0]\times 10^{-4}\nnmb~~~\cite{Barberio:2006bi}\\
\Delta a_{\mu} &\in& [-5,\,50]\times 10^{-10}~~~\cite{Yao:2006px}\nnmb
\eea
These ranges correspond approximately to the $95\%~c.l.$ allowed values,
although we have allowed a slightly larger range for $\Delta a_{\mu}$.
The constraint from the muon magnetic moment is particularly
interesting in HENS scenarios, and we shall discuss it more
extensively below.  In the immediate analysis we do not include
the LEP~II bound on the lightest Higgs boson mass.  
We will discuss this constraint below as well.

\begin{figure}[tb]
\vspace{1cm}
\centerline{
        \includegraphics[width=0.7\textwidth]{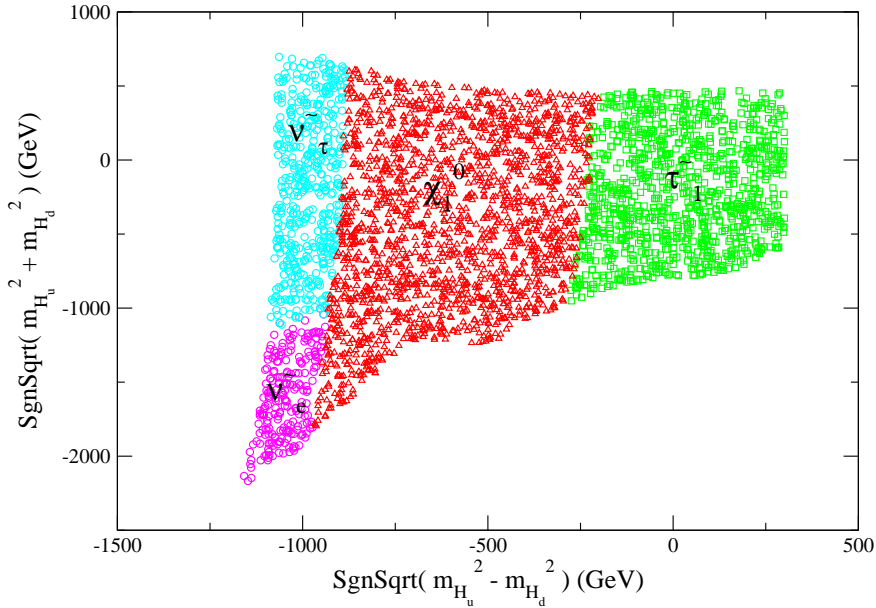}}
        \caption{Allowed parameter regions for $\tan\beta = 10$ and 
$M_{1/2} = 300\gev$.  The differently colored regions in the figure
indicate the identity of the lightest superpartner.  The quantities 
$m_{H_u}^2$ and $m_{H_d}^2$ are evaluated at the input
scale $M_{GUT}$.}
\label{lsp-10-300}        
\end{figure}

\begin{figure}[tb]
\vspace{1cm}
\centerline{
        \includegraphics[width=0.7\textwidth]{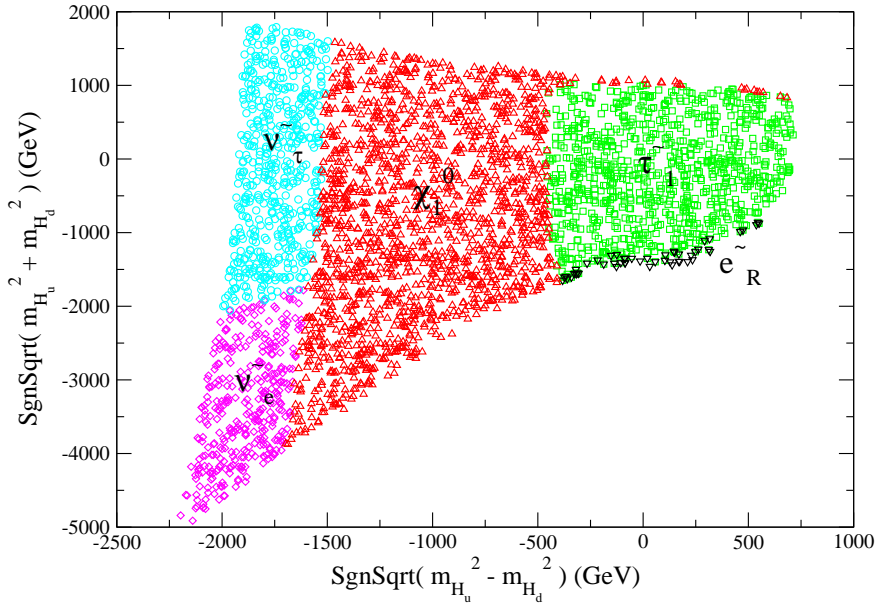}}
        \caption{Allowed parameter regions for $\tan\beta = 10$ and 
$M_{1/2} = 500\gev$.  The differently colored regions in the figure
indicate the identity of the lightest superpartner. The quantities 
$m_{H_u}^2$ and $m_{H_d}^2$ are evaluated at the input
scale $M_{GUT}$.}
\label{lsp-10-500}        
\end{figure}

\begin{figure}[tb]
\vspace{1cm}
\centerline{
        \includegraphics[width=0.7\textwidth]{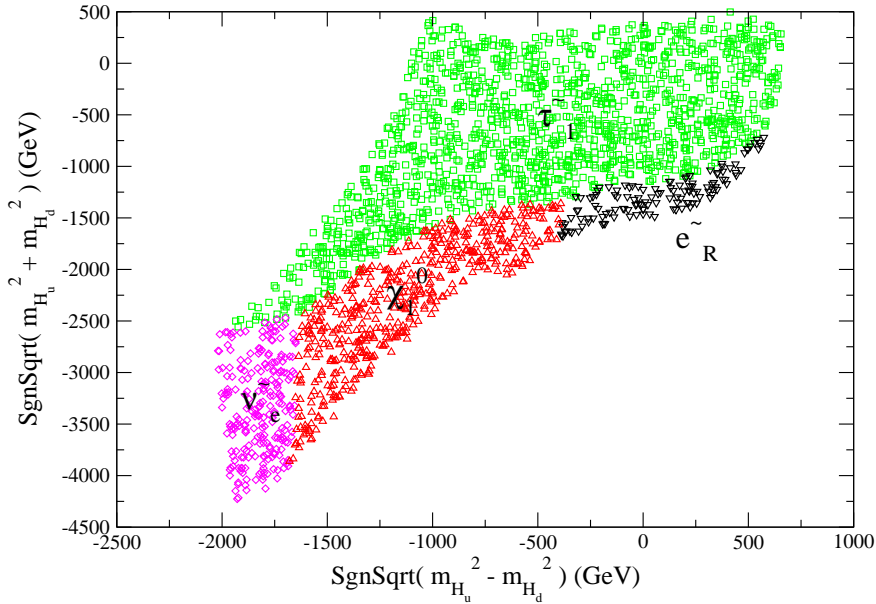}}
        \caption{Allowed parameter regions for $\tan\beta = 30$ and 
$M_{1/2} = 500\gev$.  The differently colored regions in the figure
indicate the identity of the lightest superpartner. The quantities 
$m_{H_u}^2$ and $m_{H_d}^2$ are evaluated at the input
scale $M_{GUT}$.}
\label{lsp-30-500}        
\end{figure}

  Figures~\ref{lsp-10-300}, \ref{lsp-10-500}, and \ref{lsp-30-500} 
show the allowed regions of $m_{H_u}^2(M_{GUT})$ 
and $m_{H_d}^2(M_{GUT})$, for $(\tan\beta,~M_{1/2})$ equal to $(10,\,300\gev)$,
$(10,\,500\gev)$, and $(30,\,500\gev)$, subject to the constraints described
above.  The soft Higgs masses in these plots
are re-expressed in terms of the more convenient combinations 
$SgnSqrt(m_{H_u}^2\!-\!m_{H_d}^2)(M_{GUT}) = S_{GUT}/\sqrt{|S_{GUT}|}$, 
and $SgnSqrt(m_{H_u}^2\!+\!m_{H_d}^2)(M_{GUT})$, where $SgnSqrt$ 
denotes the signed square root ($SgnSqrt(\pm x^2)=\pm x$).
Also shown in these plots is the identity of the lightest superpartner 
at each allowed parameter point.

  These figures confirm our previous approximate analysis.  
When $S_{GUT}$ is positive or zero, the LSP is a mostly right-handed
stau or selectron.  As $S_{GUT}$ becomes more negative, a neutralino
becomes the LSP, while for very large and negative values of $S_{GUT}$
the LSP is a sneutrino.  For extremely large positive or negative values
of $S_{GUT}$, one of the slepton soft masses becomes tachyonic.
The allowed parameter region is cut off
at larger positive values of $(m_{H_u}^2\!+\!m_{H_d}^2)(M_{GUT})$
because $|\mu|^2$ only has a negative solution, implying that electroweak
symmetry breaking is not possible.\footnote{In fact, the parameter
space is cut before $\mu$ reaches zero by the $BR(b\to s\gamma)$ and the 
chargino mass constraints.}  Note that in Fig.~\ref{lsp-10-500}, there is 
a thin strip along the upper border of the allowed region in
which the LSP is a neutralino.  In this strip, the $\mu$ term
is smaller than $M_1$ and the neutralino LSP is mostly Higgsino.
For larger negative values of $(m_{H_u}^2\!+\!m_{H_d}^2)(M_{GUT})$, 
$M_{A^0}^2 \to 0$ and the parameter space gets cut off by the bound
from $BR(b\to s\gamma)$.  As $(m_{H_u}^2\!+\!m_{H_d}^2)(M_{GUT})$
becomes even smaller, electroweak symmetry breaking
ceases to occur.

  The effects of the $\tau$ Yukawa coupling and left-right mixing can be
seen by comparing Figs.~\ref{lsp-10-500} and \ref{lsp-30-500}.  
In the models we are
considering, the value of the Yukawa-dependent term in 
Eq.~\eqref{2:msoftrun} for the right-handed stau soft mass is
\beq
X_{E_3} \simeq 2|y_{\tau}|^2\,m_{H_d}^2.
\eeq
The left-right mixing is also proportional to the
$\tau$ Yukawa.  As $\tan\beta$ increases, so too does the $\tau$ Yukawa,
and therefore also the Yukawa effect on the running and the mixing.
Left-right mixing tends to push the lighter stau mass lower, and 
for this reason it is more difficult to obtain a neutralino LSP
at larger values of $\tan\beta$.  However, there is also a competing
effect from the influence of the $\tau$ Yukawa on the running of
$m_{E_3}^2$.  When $m_{H_d}^2$ is large and negative, the $X_{E_3}$ term
increases the value of $m_{E_3}^2$ at low energies.
Thus, a selectron or an electron sneutrino is the LSP
in some parts of the parameter space.  

  In parts of the parameter space shown in 
Figs.~\ref{lsp-10-300}-\ref{lsp-30-500} 
the value of $m_{H_u}^2$ is large and negative,
while the slepton soft masses are considerably smaller 
in magnitude (but positive).
In these regions, it is likely that the standard MSSM minimum
is only metastable, and that a deeper charge-breaking minimum 
exists at large field values~\cite{Casas:1995pd}.  The precise constraints on 
the existence of such non-standard global minima depend on the 
details of the thermal history of the Universe~\cite{therm}, and we do not
investigate them in the present work.  However, a necessary condition
is that the lifetime of the metastable MSSM vacuum at $T=0$ should be
greater than the age of the universe. While beyond the scope of the present
work, it is possible that some regions of the parameter space shown in
Figs.~\ref{lsp-10-300}-\ref{lsp-30-500}, especially the lower left region
and at larger values of $\tan\beta$,
may not be populated after a more detailed analysis.

\subsection{Constraints from $(g-2)_{\mu}$}

  Since the sleptons in HENS models are
relatively light, the corrections to the anomalous magnetic moment
of the muon, $a_{\mu} = (g\!-\!2)_{\mu}/2$ can be 
significant~\cite{susyg-2}.  
Currently, the measured value of $a_{\mu}$ exceeds the 
SM prediction by about two standard deviations~\cite{Yao:2006px},
\beq
\Delta a_{\mu} = a_{\mu}^{exp}-a_{\mu}^{SM} = (22\pm 10)\times 10^{-10}.
\eeq
This result is suggestive of new physics.

  In the MSSM, there are additional contributions to $(g\!-\!2)_{\mu}$
from loops involving a virtual chargino and muon sneutrino, and loops 
with a virtual neutralino and smuon.  For the HENS scenarios we are studying,
in which all masses scale predominantly with $M_{1/2}$ and the gaugino
masses are universal (and assumed real and positive), the leading
supersymmetry contribution to $a_{\mu}$ is proportional to $\tan\beta$,
scales roughly as $M_{1/2}^{-2}$, and has a sign equal to the sign
of the $\mu$ term, $sgn(\mu)$~\cite{Moroi:1995yh}.  Given the  
tension between the measured value of $\Delta a_{\mu}$ and the SM prediction,
$sgn(\mu) > 0$ is strongly favored.
Indeed, we find that negative $sgn(\mu)$ is only possible for very large 
values of $M_{1/2}$.  Conversely, if $sgn(\mu)$ is positive 
the new supersymmetric contribution can help to explain this possible 
discrepancy between the SM prediction and experiment.

\begin{figure}[tb]
\vspace{1cm}
\centerline{
        \includegraphics[width=0.7\textwidth]{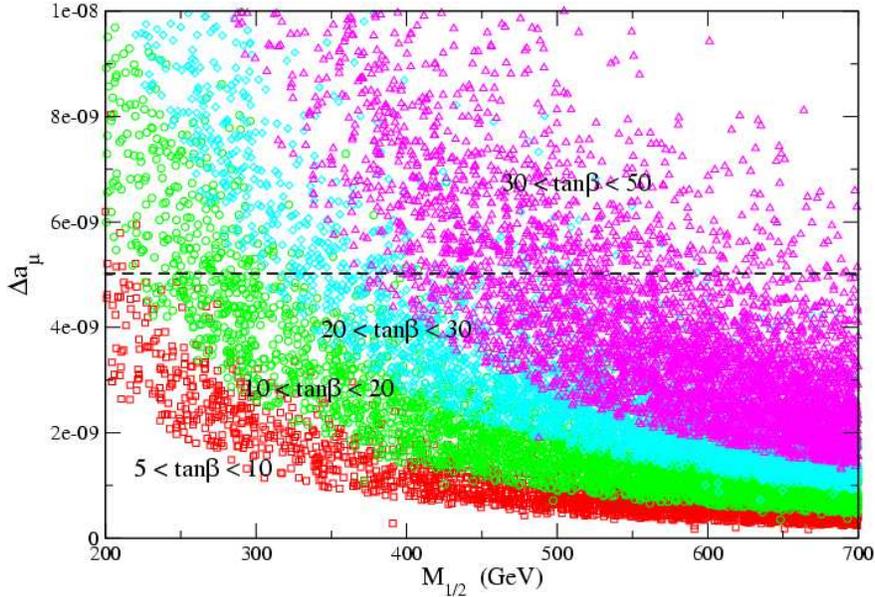}}
        \caption{$\Delta a_{\mu}^{SUSY}$ as a function of $M_{1/2}$
for several ranges of $\tan\beta$, and $sgn(\mu) > 0$.  The spread
of points come from scanning over the acceptable input values of 
$m_{H_u}^2$ and $m_{H_d}^2$.  The red points indicate $\tan\beta \in [5,10)$,
the green points $\tan\beta \in [10,20)$, the blue points 
$\tan\beta \in [20,30)$, and the magenta points $\tan\beta \in [30,50)$.
}
\label{g2-mu}        
\end{figure}

  The value of $\Delta a_{\mu}^{SUSY}$ is shown as a function 
of $M_{1/2}$ in Fig.~\ref{g2-mu}.  In generating this figure, we have taken
$sgn(\mu) > 0$, and have scanned over input values of $m_{H_u}^2$ and 
$m_{H_d}^2$ at $M_{GUT}$.
The distribution for $sgn(\mu) < 0 $ looks the same, except the sign 
of $\Delta a_\mu$ is opposite. 
With $sgn(\mu)> 0$, the new physics contribution is frequently too
large, and from this we obtain a lower bound on $M_{1/2}$ as a function
of $\tan\beta$.  For $\tan\beta = 10$, this bound is 
$M_{1/2} \gtrsim 250~\gev$, while for $\tan\beta = 30$, it increases
to $M_{1/2} \gtrsim 400~\gev$.

\subsection{Constraints from the Higgs Boson Mass}

\begin{figure}[tb]
\vspace{1cm}
\centerline{
        \includegraphics[width=0.7\textwidth]{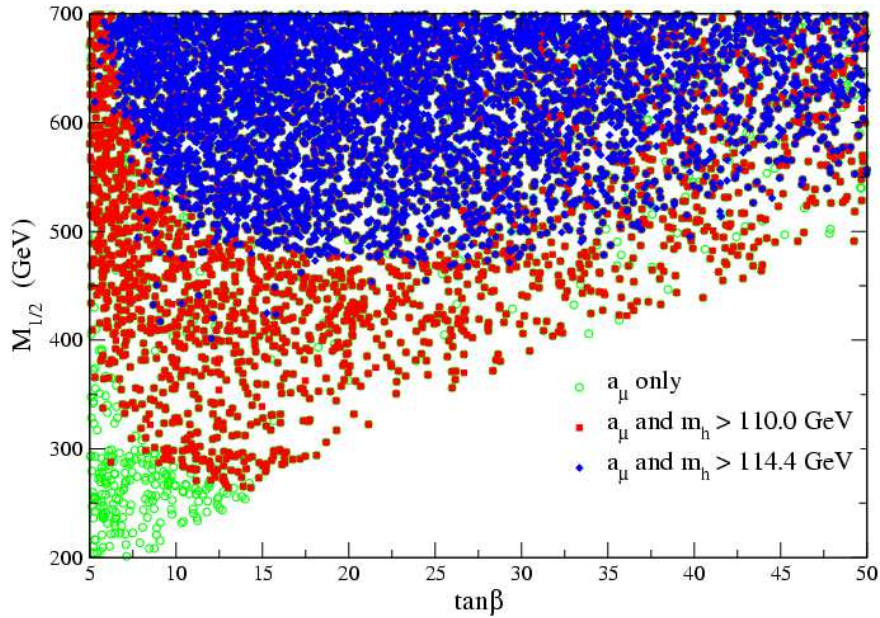}}
        \caption{Scatter plot in the $M_{1/2}-\tan\beta$ plane of 
	  solutions that respect the bounds of 
	  $\Delta a_{\mu}^{SUSY}<50\times 10^{-10}$
          and $m_h>114.4\gev$. Due to uncertainty in the top quark mass, 
	  and the theoretical uncertainty in the computation of $m_h$, 
	  a more conservative constraint on this theoretically computed 
	  value of $m_h$ is $110\gev$, which is also shown in the figure.}
\label{tanb-m12}        
\end{figure}

  A further constraint on HENS models, and one 
we have not yet imposed, is that the SM Higgs boson mass should exceed 
the LEP~II bound~\cite{Barate:2003sz},
\bea
m_h > 114.4~\gev.
\eeq
This bound also applies to the lightest CP-even Higgs boson in much
of the parameter space of the MSSM.
However at tree-level in the MSSM, the lightest CP-even Higgs boson has 
a mass below $M_Z$.  It is only because of large loop corrections to the mass, 
predominantly due to the scalar tops, that this Higgs state 
can be raised above the LEP~II bound.  With vanishing input scalar soft
masses, the stop masses scale with $M_{1/2}$.  The Higgs boson
mass bound therefore imposes a further lower bound on the universal
input gaugino mass.\footnote{One could increase the Higgs mass by introducing
a SM singlet Higgs field to the spectrum, but this introduces tensions
with grand unification~\cite{Morrissey:2005uz}.}  

  The combined bounds on $M_{1/2}$ as a function of $\tan\beta$ 
from the conditions $\Delta a_{\mu} < 50\times 10^{-10}$ and 
$m_h > 110.0~(114.4)$~GeV are shown in 
Fig.~\ref{tanb-m12}. We impose a slightly weaker $110\gev$ lower
bound on the Higgs boson mass than the $114\gev$ LEP~II bound 
to account for various uncertainties associated with the theoretical
computation of $m_h$.
We have taken $m_t = 171.4~\gev$ in our analysis.  
At smaller $\tan\beta$, less than about 15, the Higgs mass bound imposes 
the stronger constraint, while the upper bound on $\Delta a_{\mu}$ is 
more significant for values of $\tan\beta > 15$.  
For any value of $\tan\beta$, $M_{1/2}$ must be greater than about
$300$~GeV if we impose the weaker Higgs mass bound ($m_h > 110.0$~GeV), 
and larger than about $500$~GeV to satisfy the stronger bound 
($m_h > 114.4$~GeV).  Note that as $M_{1/2}$ grows, 
the phenomenological constraints on the model tend to weaken, 
but usually at the cost of increased fine-tuning in the Higgs 
sector~\cite{Barbieri:1987fn}.

\section{Dark Matter and Cosmological Signals}
\label{dark}

  In the previous section, we found that in a large region of 
the HENS parameter space a slepton or a sneutrino is the LSP.
Such an LSP can be problematic for cosmology.
If the LSP is a charged slepton, the very strong constraints that exist for 
charged stable particles imply that such a scenario is all but 
ruled out~\cite{Gould:1989gw}.  
These bounds do not apply to a sneutrino LSP, but in this case 
the direct detection rate is much too high 
(for a mass of $\sim 100~\gev$)~\cite{Falk:1994es}.
Without invoking some additional and interesting mechanisms that
allow the sneutrino as a more viable LSP~\cite{sneutrinook},
we will focus on the case of a neutralino LSP.  
Also,  a slepton as the lightest superpartner particle, charged or not,
may still provide a consistent picture of dark matter if the gravitino
is the LSP~\cite{Feng:2005ee}.  We will comment briefly on 
that possibility later.

\subsection{Neutralino LSP Relic Density}  

  We investigate the relic density of neutralino LSPs at various points in
the allowed parameter space with the aid of 
DarkSUSY~4.1~\cite{darksusy}.  This computer program 
performs a fully relativistic computation of the relic density,
and includes all relevant coannhilation channels.  

\begin{figure}[tb]
\vspace{1cm}
\centerline{
        \includegraphics[width=0.7\textwidth]{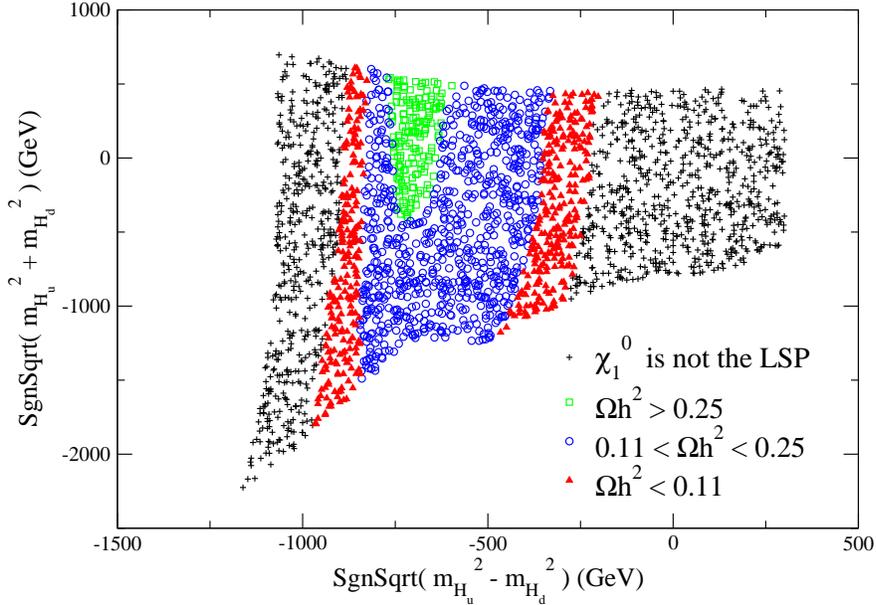}}
        \caption{Neutralino LSP relic density for $\tan\beta=10$,
$M_{1/2} = 300$~GeV, and $sgn(\mu)>0$.  The region in which the 
lightest neutralino is not the LSP is denoted by the black plus signs.
The red triangles indicate parameter points where the neutralino LSP
relic density is less than $\Omega\,h^2 < 0.11$.  In the blue and green
regions, the neutralino LSP relic density exceeds this value.}
\label{10-300}        
\end{figure}

\begin{figure}[tb]
\vspace{1cm}
\centerline{
        \includegraphics[width=0.7\textwidth]{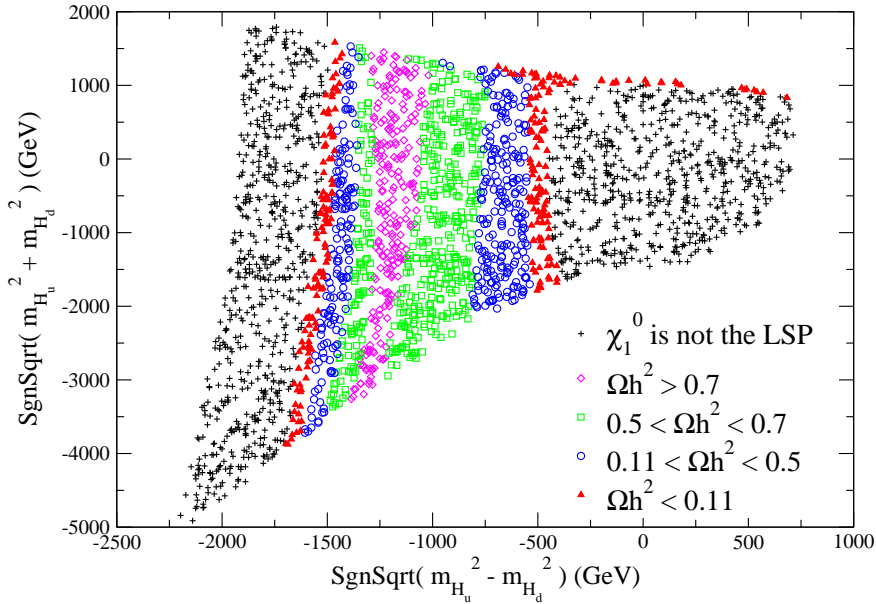}}
        \caption{Neutralino LSP relic density for $\tan\beta=10$,
$M_{1/2} = 500$~GeV, and $sgn(\mu)>0$.  The region in which the 
lightest neutralino is not the LSP is denoted by the black plus signs.
The red triangles indicate parameter points where the neutralino LSP
relic density is less than $\Omega\,h^2 < 0.11$.  In the blue, green,
and magenta regions, the neutralino LSP relic density exceeds this value.}
\label{10-500}        
\end{figure}

  Figures~\ref{10-300} and \ref{10-500} show the neutralino LSP relic
density for $\tan\beta = 10$, $M_{1/2} = 300$~GeV and $500$~GeV, 
and $\mu>0$ for the full range of allowed input values of $m_{H_u}^2$
and $m_{H_d}^2$ at $M_c=M_{GUT}$.  In both figures the black plus signs 
indicate the regions in which the lightest neutralino is not the LSP, 
but that are otherwise allowed.
The red triangles correspond to parameter points where the neutralino 
relic density is acceptably small, $\Omega\,h^2 < 0.11$.
This is to be compared with the observed dark matter 
density~\cite{Spergel:2006hy},
\beq
\Omega h^2 = 0.1045^{+0.0072}_{-0.0095}~~~~~(\mbox{WMAP only}).
\eeq
In the blue, green, and magenta regions, the neutralino relic density 
exceeds $0.11$.  These regions can still be consistent with the WMAP
measurements if there is a late-time injection of entropy into
the universe after the neutralinos have frozen out~\cite{entropy}.

  The shape of the neutralino relic density contours in 
Figs.~\ref{10-300}-\ref{10-500} can be understood 
in terms of the mass spectrum of the model.  In most of the 
parameter space, the lightest neutralino is predominantly
Bino. The main annihilation channels in this bulk region are 
$t$-channel slepton exchanges.  This is not efficient 
enough to reduce the relic density to an acceptable level
for the values of $M_{1/2}$ that are relevant, although
it does come close for $M_{1/2} = 300$~GeV.  

  Along the left and right edges of the neutralino region, the
mass of the neutralino LSP approaches that of the lightest slepton.
This near degeneracy allows for coannihilation between the neutralino
LSP and the lightest slepton to become effective, pushing the neutralino
relic density to a value well below the WMAP value.  Moving away from 
these edges towards the bulk region, the coannihilation efficiency 
falls off quickly, roughly as $\exp[-(m_{\tilde{l}}\!-\!m_{\chi^0})/T]$, 
and the neutralino density goes up.  In this transitional region,
where slepton coannihilation is only moderately efficient, the correct 
dark matter density is obtained.  
The strip of low relic density along the top of the neutralino
LSP region arises because the $\mu$ parameter becomes small.
In this strip, the neutralino LSP develops a significant Higgsino
component allowing it to annihilate effectively through gauge
bosons, and by coannihilation with the lightest chargino.

\begin{figure}[hbt]
\vspace{1cm}
\centerline{
        \includegraphics[width=0.7\textwidth]{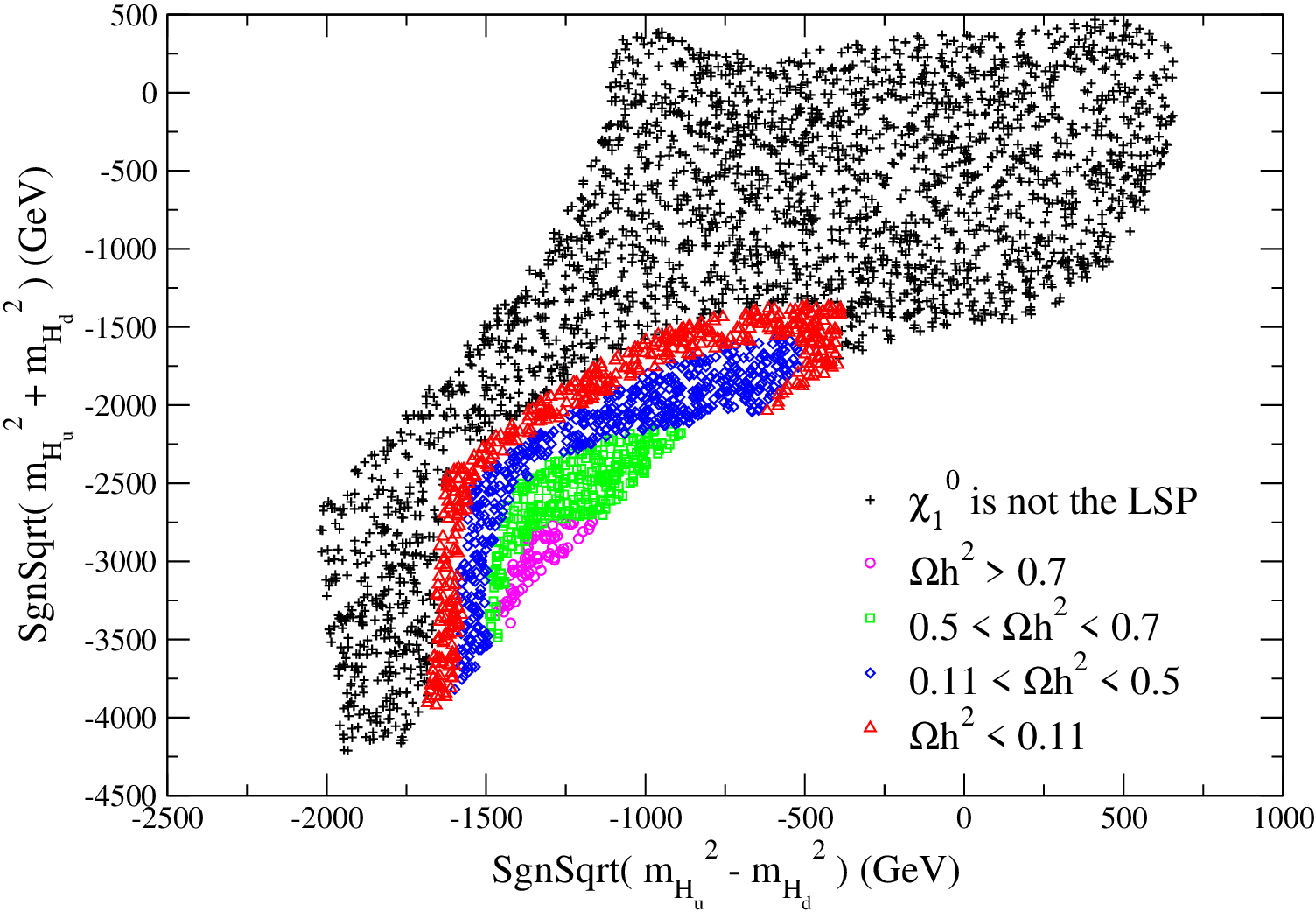}}
        \caption{Neutralino LSP relic density for $\tan\beta=30$,
$M_{1/2} = 500$~GeV, and $sgn(\mu)>0$.  The region in which the 
lightest neutralino is not the LSP is denoted by the black plus signs.
The red triangles indicate parameter points where the neutralino LSP
relic density is less than $\Omega\,h^2 < 0.11$.  In the blue, green,
and magenta regions, the neutralino LSP relic density exceeds this value.}
\label{30-500}        
\end{figure}

  In the regions where the lightest neutralino is not the 
lightest SM superpartner (denoted by black plus signs
in Figs.~\ref{10-300}-\ref{10-500}) the lightest superpartner is
always a slepton.  On the left-hand side of the neutralino region, 
this particle is either a tau or an electron sneutrino, 
as illustrated in Figs.~\ref{lsp-10-300}-\ref{lsp-30-500}.  
On the right, the lightest superpartner is a mostly right-handed 
stau or selectron.  The annihilation of sleptons is very efficient in
the early universe.  If such a particle were stable, the relic density 
in the regions discussed above would be on the order of  
$\Omega\,h^2 \sim 10^{-3}\!-\!10^{-2}$.  A relic density of heavy
charged particles of this size is firmly ruled out by 
direct searches~\cite{Gould:1989gw}.
Even for a stable sneutrino, a relic density of this size is ruled
out by dark matter direct detection searches~\cite{Akerib:2006ri}.

  A charged or neutral slepton lighter than the lightest neutralino
may still be acceptable provided the gravitino is the true LSP.
Since the gravitino couples very weakly to the MSSM states, 
the slepton NLSP would freeze out as if it were the LSP, 
and decay into gravitinos at a much later time.
The final gravitino density produced by these decays is determined 
by the quasi-stable NLSP density, $\Omega_{\tilde{l}}\,h^2$, through 
the relation~\cite{Feng:2005ee}
\beq
\Omega_{3/2}^{decay} h^2 = \frac{m_{3/2}}{m_{\tilde{l}}}\Omega_{\tilde{l}}h^2,
\eeq
where $m_{3/2}$ is the gravitino mass. 
In the parameter regions we are considering, this density is too small to 
account for all the dark matter if NLSP decays are the only source 
of relic gravitinos.  
However, there are other possible sources of relic gravitinos,
such as thermal production after inflationary reheating~\cite{Ellis:1982yb}
and non-thermal production through heavy particle decays~\cite{Dine:2006ii} 
that can bring the total gravitino relic density up to the value needed to 
explain all the dark matter.  Let us also note that this scenario
is constrained by the requirement that the late-time decays
not overly disrupt the predictions of big-bang nucleosynthesis or
the black-body spectrum of the cosmic microwave background radiation.
These constraints are relatively weak for a sneutrino NLSP, but they
are quite severe for a charged slepton (or neutralino) NLSP,
and require $m_{3/2} \lesssim 100$~MeV for most of the range 
of NLSP masses we are considering~\cite{Feng:2004zu}.

  We have also examined the effect of varying $\tan\beta$ on the
predictions for dark matter within the model.  Smaller values
of this ratio, $\tan\beta < 10$, do not change 
the qualitative features of the picture described above.
For $\tan\beta > 10$, there are a couple of important changes.
Most importantly, larger values of $\tan\beta$ induce more mixing 
between the left- and right-handed staus, which has the effect of lowering
the mass of the lightest stau.  Because of this, for $\tan\beta = 30$
and $M_{1/2} = 500$~GeV there is no longer an acceptable region of 
parameter space in which the lightest neutralino is both the 
lightest superpartner, and mostly Higgsino.  In Fig.~\ref{30-500}  
we plot the neutralino LSP relic density  for $\tan\beta=30$,
$M_{1/2}=500\gev$, and $sgn(\mu)>0$.  The thin slice of acceptable
$\Omega h^2$ arises due to the coannihilation of the LSP with a
light slepton.

\subsection{Direct and Indirect Detection of Dark Matter}

  The prospects for direct and indirect detection of neutralino dark
matter were investigated using DarkSUSY~4.1~\cite{darksusy}.
For the most part, the direct and indirect detection signals
within our scenario are very similar to those of a generic
mSUGRA model with a mostly Bino LSP~\cite{Baer:2004qq}.  
In general, these potential signals are very weak.  
However, much stronger signals can arise when the neutralino
LSP has a significant Higgsino component.

\begin{figure}[tb]
\vspace{1cm}
\centerline{
        \includegraphics[width=0.7\textwidth]{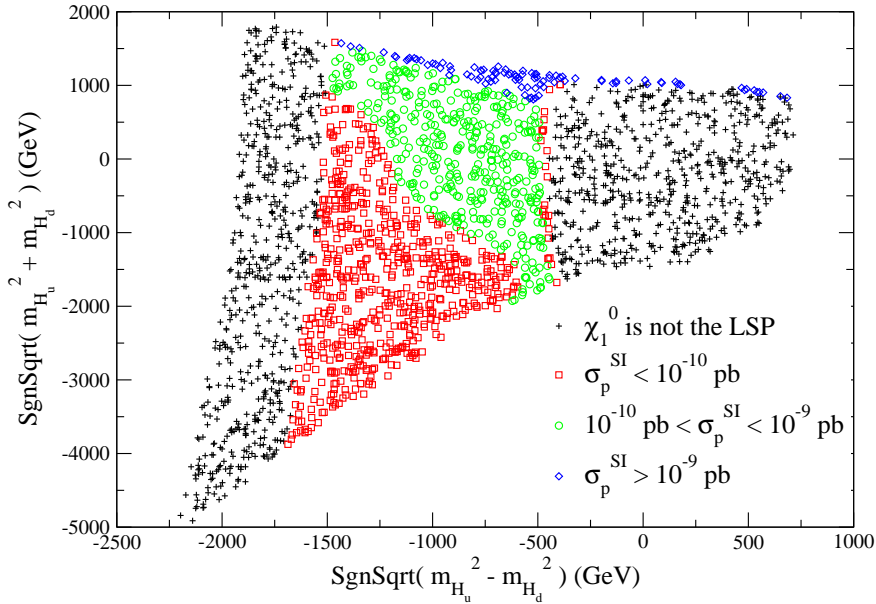}}
        \caption{
Neutralino-nucleon spin-independent elastic scattering
cross section for $\tan\beta = 10$ and $M_{1/2} = 500$~GeV.
The black plus signs indicate the parameter region that is
consistent with various phenomenological constraints, but need
not have a neutralino LSP.  The red, green, blue, magenta points
indicate parameter points for which a neutralino is the LSP,
for several ranges of $\sigma_{p}^{SI}$.}
\label{dd-10-500}        
\end{figure}

  Dark matter in the local halo can be detected directly by its 
elastic scattering with heavy nuclei.  The most strongly constrained 
neutralino-nucleus cross-sections are the spin-independent ones.
It is conventional to express the experimental limits on these 
cross-sections in terms of an effective neutralino-proton cross-section.  
The values of the effective spin-independent cross-sections 
in the present model, for $\tan\beta = 10$ and $M_{1/2}$ = 500~GeV,
are shown in Fig.~\ref{dd-10-500}.\footnote{In these figures, the effective
cross-section is rescaled by the relic density of the neutralino LSP
when it is less than the WMAP value.  Such a rescaling is conventional,
and accounts for the fact that a smaller density of neutralinos will
lead to fewer events in an experiment.  This rescaling is also the
reason why there is a thin strip with a low effective cross-section
on the right-hand side of the allowed region in Fig.~\ref{dd-10-500}.}
Except in a thin strip at the top of the allowed region, 
where the neutralino LSP is predominantly Higgsino, these cross-sections
are much smaller than the current experimental bound 
(for a standard set of assumptions about the local halo density) 
of $\sigma_p^{SI} < 10^{-6}\!-\!10^{-7}$~pb~\cite{Akerib:2006ri} for the
range of neutralino masses we consider here.  The scattering cross-sections
in the Higgsino region are within an order of magnitude this bound.

\begin{figure}[tb]
\vspace{1cm}
\centerline{
        \includegraphics[width=0.7\textwidth]{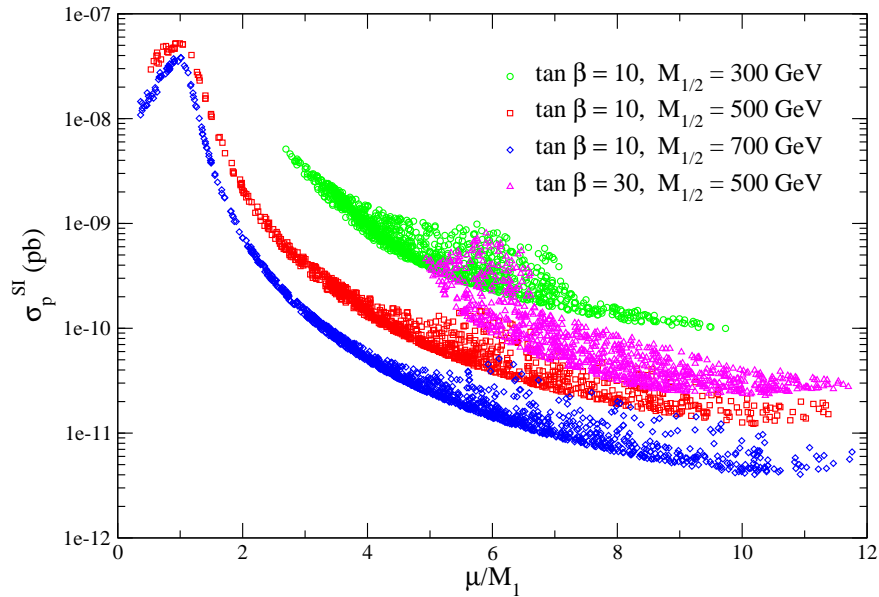}}
        \caption{Neutralino-nucleon elastic scattering cross-section
for $\tan\beta = 10$ and $M_{1/2} = 300$,~500, and 700~GeV, 
as well as for $\tan\beta = 30$ and $M_{1/2} = 500\,\gev$, 
as a function of the ratio $\mu/M_1$.  The current experimental limit for
standard assumptions about the local halo density is
$\sigma_p^{SI} < 10^{-6}\!-\!10^{-7}$~pb for the
range of neutralino masses we consider here~\cite{Akerib:2006ri}.
}
\label{dd-mu-m1}
\end{figure}

  Spin-independent scattering between a neutralino and a nucleon 
is mediated predominantly by the exchange of CP-even Higgs 
bosons and squarks.  In the allowed 
parameter regions discussed above, the squarks tend to be much 
heavier than the neutralino LSP, thereby suppressing their contribution.  
Thus, the leading contribution to the spin-independent neutralino-nucleon 
cross-section usually comes from CP-even Higgs exchange.  Since the 
neutralino Higgs vertices are proportional to the Higgsino 
component of the neutralino, the effective cross-section depends 
sensitively on the ratio $\mu/M_1$.  This is illustrated in 
Fig.~\ref{dd-mu-m1} for $\tan\beta = 10$ and $M_{1/2}= 300,~500,~700$~GeV, 
as well as for $\tan\beta = 30$ and $M_{1/2} = 500\,\gev$.  
The dependence on other parameters such as $M_{A^0}$ and 
$m_{\tilde{q}}$ is much weaker.

\begin{figure}[tb]
\vspace{1cm}
\centerline{
        \includegraphics[width=0.7\textwidth]{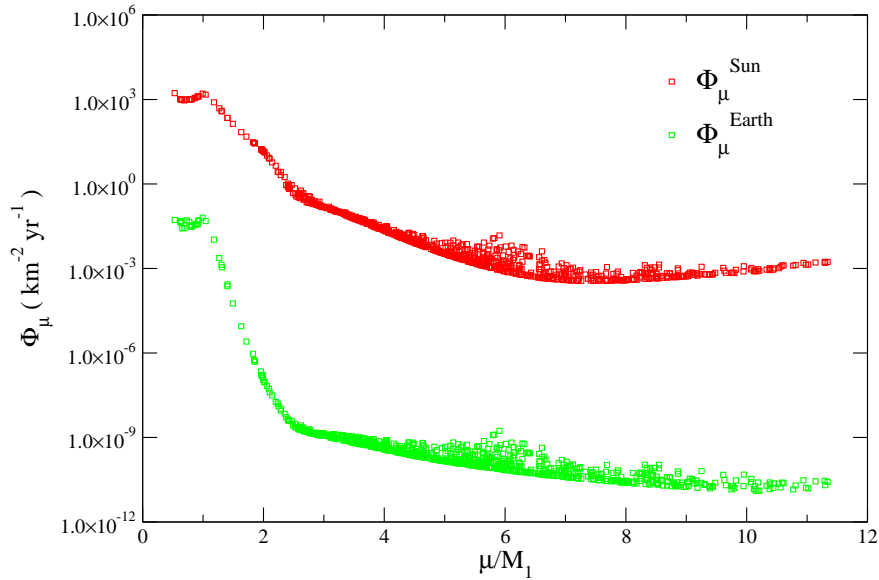}}
        \caption{Net muon flux induced by neutralino dark matter
annihilation in the core of the sun~(red circles) and the Earth~(green
squares) for $\tan\beta = 10$
and $M_{1/2} = 500$~GeV.  A detector threshold of 1\;GeV is assumed. The best
current bound is $\Phi_{\mu} \lesssim 10^3\,km^{-2}\,yr^{-1}$ coming from
Super-Kamiokande~\cite{Desai:2004pq}.  Future sensitivities from IceCube
and ANTARES are $\Phi_{\mu} = 10^1-10^2\,km^{-2}\,yr^{-1}$. 
}
\label{muflux}
\end{figure}

  Relic neutralinos can also be searched for indirectly 
by looking for their annihilation products in regions
where they tend to clump, and have a local density much larger
than the average value.  In these particularly dense patches, 
the rate of neutralino annihilation can become large enough that 
their products are potentially observable.
The most stringent of these indirect detection constraints on 
neutralino dark matter usually comes from their annihilation in the core
of the sun and the earth.\footnote{The rates for other indirect
dark matter signals, such as from positron emission, seem to be quite small
except when the neutralino LSP is mostly Higgsino, or if the annihilation
cross section is enhanced by a $s$-channel CP-odd Higgs resonance.}
The neutrinos produced by this process can lead to a signal, 
in the form of a muon flux, in neutrino telescopes.  

  The best current bound on such a muon flux comes from the Super-Kamiokande 
experiment~\cite{Desai:2004pq}, and is on the order of 
$\Phi_{\mu} \lesssim 10^3\,km^{-2}\,yr^{-1}$.
This sensitivity or bound is expected to be tightened to 
$10^1-10^2\,km^{-2}\,yr^{-1}$ 
in the next few years by IceCube~\cite{Toale:2006ti} and 
ANTARES~\cite{Hossl:2004ay}.
The values of $\Phi_{\mu}$ due to annihilation in both the core of
the earth and the sun in the present model with $\tan\beta = 10$
and $M_{1/2} = 500$~GeV are shown in Fig.~\ref{muflux}
as a function of the ratio $\mu/M_1$, assuming a detector threshold
of $1\;\gev$.  As for the direct detection signal,
the largest indirect detection signal is obtained when $\mu/M_1$ is 
on the order of unity and the neutralino LSP has a significant 
Higgsino component.  In fact, for $\mu/M_1\sim 1$, 
the signal as at or slightly above the Super-Kamiokande bound.  
For larger values of $\mu/M_1$ the signal falls off very quickly
to values that are well below the reach of upcoming experiments.

\section{Collider Signatures}
\label{coll}

  In HENS scenarios, the sleptons and the electroweak gauginos are 
generally very light relative to the squarks and the gluino.  
If the lightest neutralino is the LSP, which we assume
throughout this section, the distinguishing feature of these scenarios 
at colliders are multi-lepton events with missing $E_T$.  In this section
we discuss the prospects for discovery and identification of HENS models
at the Tevatron and the LHC.

\subsection{Trilepton Signature at the Tevatron}

  The most promising search channel at the Tevatron is the trilepton
signal with missing $E_T$~\cite{trileptons,matchev3l,baer3l}.  
This can be induced, for example, 
by the electroweak production of $\chi^0_2\,\chi_1^{\pm}$, 
with subsequent cascades of the form $\chi_2^0\to 
\tilde{\ell}_L^*\,\ell^- \to \chi_1^0\,\ell^+\,\ell^+$ and 
$\chi_1^+\to \tilde{\nu}_{\ell'}\,{\ell'}^+\to 
\chi_1^0\,\nu_{\ell'}\,{\ell'}^+$.
For $M_{1/2} \geq 300~\gev$, the significant source of SUSY events 
comes from the electroweak production of gauginos, making this channel
a copious and clean one.  In HENS scenarios, the $\chi_2^0$ and $\chi_1^{\pm}$
states tend to be mostly Wino and have two-body decays into
left-handed sleptons.  Because of this feature, the branching fractions 
of the abovementioned decay cascades can be significant, leading to 
a sizeable trilepton cross-section.  Indeed, the mass spectrum derived
from HENS models is close to being optimal for trilepton production.

  To estimate the effective Tevatron trilepton cross-sections, 
we have simulated SUSY production from 
$p\bar{p}$ collisions at $\sqrt{s} = 1.96$~TeV using
ISAJET~7.74~\cite{isajet}.  Following the treatment 
in Ref.~\cite{baer3l}, we use the ISAJET subroutines CALSIM and CALINI 
(in the ISAPLT package) as a simple detector model with coverage in 
the range $-4 < \eta < 4$, and calorimeter cells of size 
$\Delta\eta\times\Delta\phi = 0.1\times 0.262$.  
To simulate energy resolution uncertainties, the electromagnetic
calorimeter cells are smeared by an amount $0.15/\sqrt{E/{\rm GeV}}$,
while the hadronic calorimeter cells are smeared by an amount
$0.7/\sqrt{E/{\rm GeV}}$.  We define jets as hadronic clusters with 
$E_T > 15$~GeV within a cone of size $\Delta R = 0.7$, 
and use the GETJET subroutine to perform the clustering.  
Isolated leptons are defined to be $e$'s or $\mu$'s having $p_T > 5$~GeV, 
with net visible hadronic activity $E_T < 2$~GeV within a cone of 
size $\Delta R = 0.4$ about the lepton direction.

  We focus on a particular set of cuts, corresponding to the 
HC2 set in Ref.~\cite{baer3l}, that is well-suited to 
the HENS mass spectrum~\cite{matchev3l,Baer:2001ze}.
In each event, we require three isolated leptons with 
$p_T(\ell_{1,2,3}) > 20,\,15,\,10~\gev$,
and $|\eta(\ell_{1,2,3})| < 2.5$.  In addition to this,
the total missing $E_T$ must exceed $25$~GeV, the invariant mass
of same-flavor opposite-sign dileptons must lie in the range
$12~\mbox{GeV} < m_{\ell\ell} < 81~\gev$, and the transverse invariant
mass between each lepton and the missing $E_T$ vector must lie outside
the range $60~\gev < m_T(\ell,\met) <85~\gev$.  The dilepton
invariant mass veto is designed to remove background events from 
off-shell $Z$ and $\gamma$ decays, while the $m_T$ veto removes
leptons from $W$ decays.  With these cuts, the  SM background
is estimated to be $0.49~fb$~\cite{baer3l}, and is due mostly to the remaining 
$W^*Z^*$ and $W^*\gamma^*$ events in which both off-shell gauge bosons decay
leptonically.

\begin{figure}[tb]
\vspace{1cm}
\centerline{
        \includegraphics[width=0.7\textwidth]{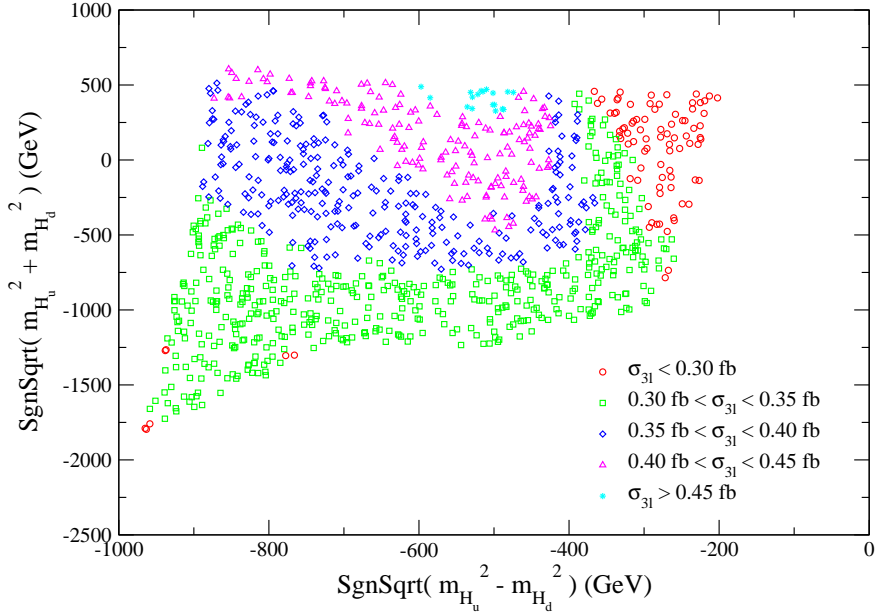}}
\vspace{1cm}
        \caption{Trilepton cross-sections after HC2 cuts at the Tevatron
for $M_{1/2}=300\;\gev$ and $\tan\beta = 10$.  The estimated background
is $0.49\,fb$~\cite{baer3l}.  }
\label{tevscan}
\end{figure}

  In Fig.~\ref{tevscan} we show the trilepton cross-section subject to
the cuts for $M_{1/2} = 300$~GeV and $\tan\beta = 10$.
This value of $M_{1/2}$ is about as small as possible within the model
given the lower bound on the light Higgs boson mass, and represents
a best-case scenario at the Tevatron.  
Note that for considerably larger values of $\tan\beta$, the constraint
from the anomalous magnetic moment of the muon requires larger values
of $M_{1/2}$ as well, as can be seen in Fig.~\ref{tanb-m12}.  
Like in the previous plots, the values of $m_{H_u}^2$ and $m_{H_d}^2$ 
in Fig.~\ref{tevscan} are those at the input scale, $M_c = M_{GUT}$.

  The dependence of the trilepton cross-section on $m_{H_u}^2$ 
and $m_{H_d}^2$ can be understood in terms of the mass spectrum.  
Except in the upper-right portion of the allowed parameter space,
the effective cross-section increases smoothly from bottom to top
as the value of $\mu$ decreases.  In most of the parameter space,
$\mu$ is larger than $M_2$ and the $\chi_2^0$ and $\chi^{\pm}_1$ states are
mostly Wino.  As $\mu$ approaches $M_2$, these states develop 
a larger Higgsino fraction and their masses are reduced by the mixing.  
The heavier chargino and neutralino states become lighter as well.  
On account of these effects, the total gaugino cross section 
is increased leading to more trilepton events.  This pattern is 
broken in the upper right corner of the parameter space because 
the mass of the $\chi_2^0$ state approaches the left-handed
slepton masses from above, again due to Higgsino mixing.  When this 
mass difference becomes small, the branching fraction for 
$\chi^2_0 \to \tilde{\ell}_L\ell$ goes down.  The leptons produced
by the cascades become relatively soft as well.  Since this decay 
mode plays a prominent role in the trilepton signal subject to 
the HC2 cuts, the effective cross-section falls off rapidly when the 
decay fraction is suppressed.  The effective cross-section in this
region can be increased by using slightly weaker lepton $p_T$ cuts,
such as the SC2 set discussed in Ref.~\cite{baer3l}, but at the 
expense of an increase in background.

  The effective trilepton cross-sections shown in Fig.~\ref{tevscan} 
fall within the range of $0.2\!-\!0.5\;fb$.  Given the estimated background 
of $0.49\;fb$, the signal significance level is  marginal.  For example,
the Poisson probability $P_p$ for a total of ten events, corresponding to 
the maximal expected signal and background with $10\,fb^{-1}$, 
is about $P_p = 0.016$. While this is unfortunately not enough for 
a discovery, an excess of clean trilepton events at the Tevatron 
would provide a tantalizing hint of a light HENS scenario.
We also note that other event signatures involving leptons can
be searched for in these scenarios.  Of particular noteworthiness is
the same-sign dilepton signature, which has small standard model
background.

\subsection{Signals at the LHC}

  If nature is supersymmetric and has a HENS spectrum, the prospects
for discovery at the LHC with $10\;fb^{-1}$ of data are 
excellent provided $M_{1/2}$ is less than about 700~GeV.  
To quantify this, we focus on six inclusive LHC SUSY search channels,
classified by the number of isolated leptons in the event:
$0\ell+\met+jets$; $1\ell+\met+jets$; $2\ell\;OS + \met + jets$;
$2\ell\;SS + \met + jets$; $3\ell + \met + jets$; $\geq 4\ell 
+ \met + jets$~\cite{Baer:1995nq,Baer:1995va}.
(Here, $OS$ and $SS$ refer to opposite-sign and same-sign dileptons, 
respectively.)  Besides an excess of events in these channels,
which is expected in many SUSY scenarios, the relative numbers of
events within different channels can point towards small input
scalar soft masses.

  Supersymmetric events at the LHC were simulated 
using ISAJET~7.74~\cite{isajet}.  We use the ISAJET subroutines 
CALSIM and CALINI (in the ISAPLT package) as a simple detector 
model with coverage in the range $|\eta| < 5$, 
and calorimeter cells of size $\Delta\eta\times\Delta\phi = 0.05\times 0.05$.  
A gaussian smearing of the calorimeter cells is included to simulate 
energy resolution uncertainties.
The electromagnetic calorimeter cells are smeared by an amount 
$0.1/\sqrt{E/{\rm GeV}}\oplus 0.01$, where $\oplus$ denotes addition
in quadrature.  Hadronic calorimeter cells are smeared by an amount
$0.5/\sqrt{E/{\rm GeV}}\oplus 0.03$ for $|\eta|< 3$, and 
$1.0/\sqrt{E/\gev}\oplus0.07$ for $|\eta|> 3$.
We define jets as clusters with $E_T > 100$~GeV and $|\eta|< 3$ within a 
cone of size $\Delta R = 0.7$, and use the GETJET subroutine
to perform the clustering.  Isolated leptons are defined to be
$e$'s or $\mu$'s having $p_T > 10$~GeV and $|\eta| < 2.5$, 
with total visible activity $E_T < 5$~GeV within a cone 
of size $\Delta R = 0.3$ about the lepton direction.

  For all channels studied, we choose a cut energy $E_T^c = 200~\gev$ 
and demand that each event have at least two hard jets, $n_j \geq 2$, 
with $E_T > E_T^c$, as well as missing transverse energy $\met > E_T^c$.  
This cut substantially reduces the SM backgrounds relative 
to the SUSY signals.\footnote{A larger value of $E_T^c$ would be preferable 
to reduce the large SM backgrounds in the $0\,\ell$ channel.
However, the $0\ell$ SUSY signal is still easily distinguishable from 
the large background for the parameter points we consider here.}  
We also require that the transverse sphericity of the event satisfies 
$S_T > 0.2$ to reduce the dijet background~\cite{atlastdr}.  
In zero-lepton events, we demand that the transverse angle
between the missing momentum vector and the nearest jet must lie
in the range $30^\circ < \Delta\phi(\met,j) < 90^\circ$.
In the one-lepton channel, we require a single isolated lepton
with $p_T > 20~\gev$, as well as $M_T(\ell,\met) > 100~\gev$ to reduce
the leptonic $W$ background.  For events with two or more isolated leptons, 
we demand that $p_T(\ell_{1,2}) > 20~\gev$ for the two hardest leptons.

\begin{figure}[tb]
\vspace{1cm}
\centerline{
        \includegraphics[width=0.7\textwidth]{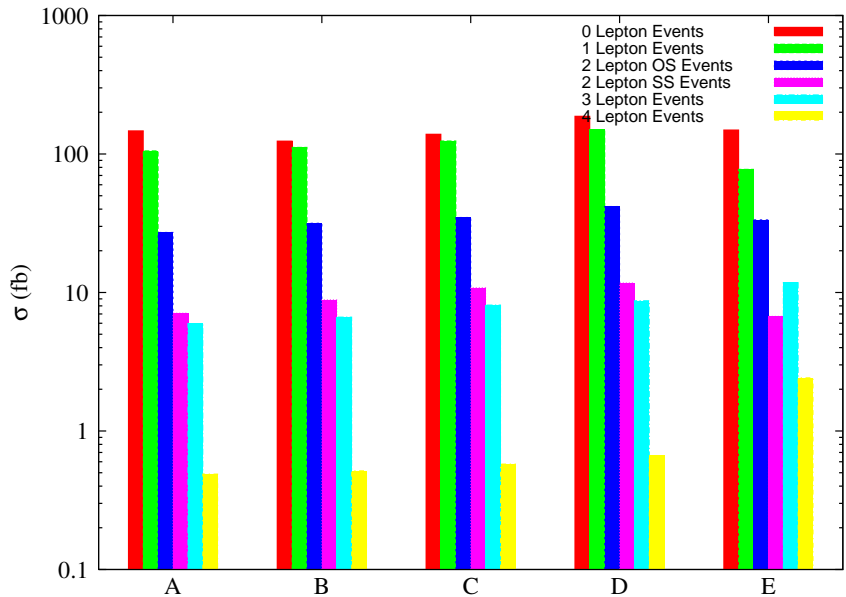}}
\vspace{1cm}
        \caption{Inclusive signal cross-sections after cuts at the LHC
for $M_{1/2}=500\;\gev$ and $\tan\beta = 10$ for the five sample points
described in the text.  For comparison, the SM backgrounds are estimated 
to be about $400~fb$, $26~fb$, $9~fb$, $0.25~fb$,  $0.1~fb$, and $0.002~fb$
for the $0\ell$, $1\ell$, $2\ell\;OS$, $2\ell\;SS$, $3\ell$ and $4\ell$ 
channels respectively~\cite{Baer:1995nq,Baer:1995va}.}
\label{lhcbar}
\end{figure}

\begin{figure}[tb]
\vspace{1cm}
\centerline{
        \includegraphics[width=0.7\textwidth]{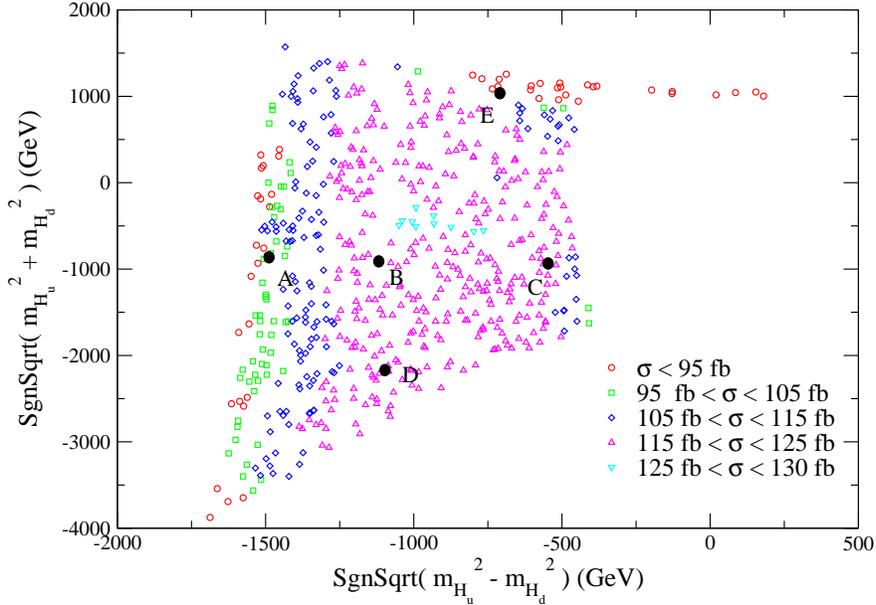}}
\vspace{1cm}
        \caption{$1\ell$ cross-sections after cuts at the LHC
for $M_{1/2}=500\;\gev$ and $\tan\beta = 10$.  The estimated background
is $26\,fb$.}
\label{10-500-scan-1l}
\end{figure}

  The cross sections after cuts for $M_{1/2} = 500\;\gev$ 
and $\tan\beta = 10$ are given in Fig.~\ref{lhcbar} for five sample points.  
For comparison, the SM backgrounds are estimated to be about
$400\,fb$, $26\,fb$, $9\,fb$, $0.25\,fb$, $0.1\,fb$, and $0.002\,fb$ for the 
$0\ell$, $1\ell$, $2\ell\;OS$, $2\ell\;SS$, $3\ell$, and $4\ell$  
channels respectively~\cite{Baer:1995nq,Baer:1995va}.
The locations of the sample points, $A,\,B,\,C,\,D,$ and $E$, 
in the $m_{H_u}^2(M_{GUT})\!-\!m_{H_d}^2(M_{GUT})$ plane are 
listed in the Appendix, and are also indicated in 
Figs.~\ref{10-500-scan-3l} and \ref{10-500-scan-1l}.
Among the five sample points, $A$ and $C$ have a neutralino relic 
density within the WMAP allowed range, while the other points lead 
to relic densities that are too large (but could be acceptable with 
a non-standard cosmology).  At point E, the $\mu$ term is 
on the same order as $M_2 \simeq 2\,M_1$, but it is greater 
than $750\,\gev$ at the other sample points.  Thus, except for point
E, the LSP is a mostly Bino neutralino, while the lightest chargino
and the next-to-lightest neutralino are predominantly Wino.  
At point E where $\mu < M_2$, all the chargino and neutralino 
states are fairly light and have significant Higgsino components,
which, as we shall discuss below, is the reason for the increase 
in the $3\ell$ and $4\ell$ rates.

\begin{figure}[tb]
\vspace{1cm}
\centerline{
        \includegraphics[width=0.7\textwidth]{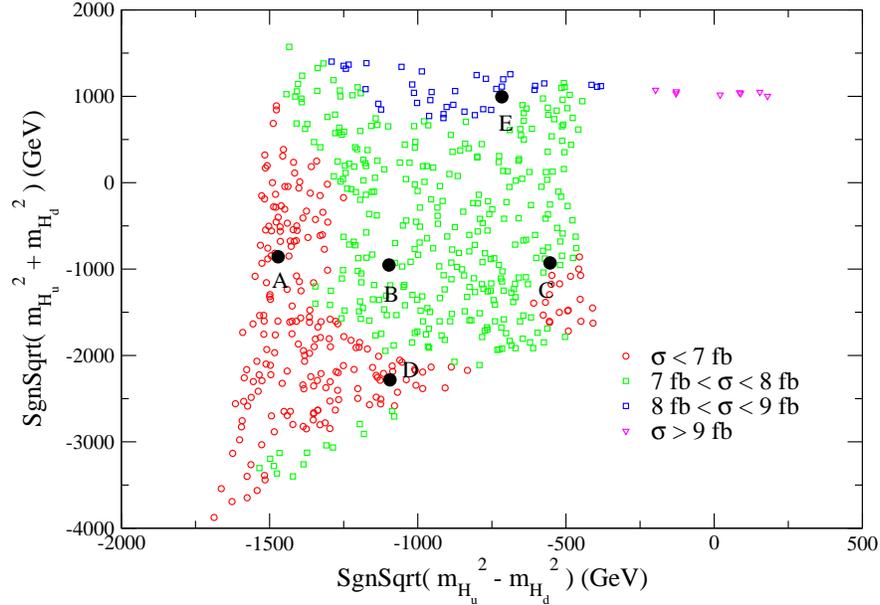}}
\vspace{1cm}
        \caption{$3\ell$ cross-sections after cuts at the LHC
for $M_{1/2}=500\;\gev$ and $\tan\beta = 10$.  The estimated background
is $0.1\,fb$.}
\label{10-500-scan-3l}
\end{figure}
 
  In each of the six search channels, the SUSY signal is easily 
distinguishable from the background with $10fb^{-1}$ of luminosity at the LHC.
A more challenging task beyond an initial discovery is to distinguish
this class of models from other (SUSY) scenarios and to deduce the
model parameters.  The number of leptonic events relative to 
the number of $0\,\ell$ events is useful in this regard.
Compared to a generic mSUGRA input spectrum with $m_0 > 0$, the 
ratio of $1\,\ell$ events to $0\,\ell$ events is much larger
for a given value of $M_{1/2}$.  For example, in mSUGRA with 
$(m_0,\,M_{1/2},\,A_0,\,\tan\beta,\,sgn(\mu)) = (200\,\gev,\;500\gev,\,0,
10,+)$, the ratio of $0\;\ell$ to $1\,\ell$ events is greater than four,
whereas this ratio is close to unity for all five sample points
considered.  The ratio of the number of $0\,\ell$ events to 
the number of $2\,\ell$ events is also much larger in a generic 
mSUGRA framework than it is here.  This is the consequence of 
having left-handed sleptons lighter than the $\chi_2^0$ and $\chi_1^{\pm}$
states, which are in turn light enough to be generated by squark decays.
For example, cascade chains such as $\tilde{q}\to {\chi}_2^0\,q \to
\tilde{\ell}_L^*\,\ell^-\,q \to \chi_1^0\,\ell^+\,\ell^-\,q$ have a 
significant branching probability, and are a rich source of leptons.

  The dependence of the effective $3\ell$ cross-section  
on the input Higgs soft mass parameters is shown in Fig.~\ref{10-500-scan-3l}.
For the most part, this dependence is fairly mild except in the 
upper right portion of the allowed region.  
Here, the $\mu$ term approaches the Bino mass $M_1$, and in the thin tail 
extending to the right, $\mu$ even falls below $M_1$.  In this region, 
all the neutralinos and charginos are significantly lighter than the squarks
and gluinos.  As a result, the decay cascades initiated by the strong 
superpartners are frequently very long, involving several chargino and 
neutralino states.  At each step in the cascade chain there is a chance
of producing a lepton, and thus the total fraction of events containing
multiple leptons is very high.  For example, the decay chain
$\tilde{u}_R \to \chi_3^0\,u \to 
\tilde{\ell}_R^*\,\ell^-\,u
\to \chi_1^0\,\ell^+\,\ell^-\,u$
is kinematically allowed when $\mu$ is small, 
and has a significant branching fraction.
The preponderance of leptons can be so high that
the number of $0\ell$ and $1\ell$ events (and even to some extent some 
$2\ell$ events) are significantly suppressed.  This can be seen in 
Fig.~\ref{10-500-scan-1l}, which shows the $1\ell$ effective cross 
section for $M_{1/2} = 500\,\gev$ and $\tan\beta = 10$.  Note that the 
small $\mu$ region is strongly constrained by direct and indirect 
searches for dark matter, and will be probed by upcoming experiments, 
as was discussed above.

  In the leftmost portion of the allowed region of 
Fig.~\ref{10-500-scan-3l}, there is also a net decrease in the cross section, 
which occurs in the other leptonic channels as well.
Within this region, leptons typically originate from decays of the 
mostly Wino $\chi_2^0$ and $\chi_1^{\pm}$ states into left-handed 
sleptons and sneutrinos, which subsequently decay into the neutralino LSP.  
However, these left-handed states are only slightly heavier than the LSP, 
so the lepton emitted from the slepton decays tends to be soft, 
making it less likely to pass the lepton $p_T$ cuts.  

  A particularly distinctive signature of the HENS models
are inclusive $4\ell$ events.  We find effective cross-sections above 
$0.5\,fb$ for $\tan\beta = 10$ and $M_{1/2} = 500\,\gev$, which is 
sufficient for a $10\,fb^{-1}$ LHC discovery given the SM background 
of about $0.002~fb$~\cite{Baer:1995va}.  There is a sharp increase 
in the $4\ell$ cross-section in the small $\mu$ region at the upper right of
the parameter space.  The dominant sources of this increase are 
cascades initiated by right-handed squarks of the type described
previously.  Furthermore, because the left-handed sleptons are
lighter than $\chi_3^0$ but heavier than $\chi_2^{\pm}$ and 
$\chi_4^0$ in this region, superpartner cascades such as 
$\tilde{u}_L \to \chi_2^{\pm} \to \tilde{\nu} 
\to \chi_3^0 \to \tilde{\ell}_R \to \chi_1^0$ accompanied by many leptons
have a non-trivial branching fraction and can produce three leptons 
from the single squark parent.\footnote{The final lepton in this 
cascade tends to be quite soft because the mass difference 
$(m_{\tilde{l}_R} - m_{\chi_1^0})$ is very small in this part of the 
parameter space.  However, the other two leptons in the cascade 
tend to be quite hard.}  As a result, $4\ell$ rates greater than 
$5\,fb$ can occur.  We have also investigated
the exclusive clean trilepton channel.  It does not appear to be
as promising as the inclusive channels.

\begin{figure}[tb]
\vspace{1cm}
\centerline{
        \includegraphics[width=0.7\textwidth]{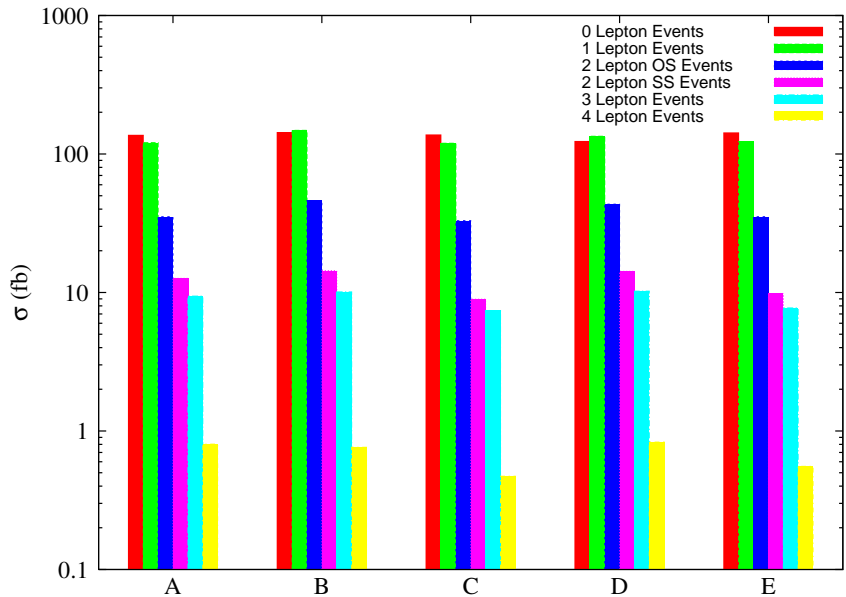}}
\vspace{1cm}
        \caption{Inclusive signal cross-sections after cuts at the LHC
for $M_{1/2}=500\;\gev$ and $\tan\beta = 30$ for the five sample points
described in the text. For comparison, the SM backgrounds are estimated 
to be about $400~fb$, $26~fb$, $9~fb$, $0.25~fb$,  $0.1~fb$, and $0.002~fb$
for the $0\ell$, $1\ell$, $2\ell\;OS$, $2\ell\;SS$, $3\ell$ and $4\ell$ 
channels respectively~\cite{Baer:1995nq,Baer:1995va}.}
\label{lhcbar30}
\end{figure}

   Varying $\tan\beta$ does not qualitatively 
affect our findings.  The cross sections after cuts for 
$M_{1/2} = 500\;\gev$ and $\tan\beta = 30$ are given in 
Fig.~\ref{lhcbar30} for five sample points, $A',\,B',\,
C',\,D'\, ,E'$.  Details of these sample points are given in the Appendix.
The cross sections in all six channels are similar to those in
Fig.~\ref{lhcbar} with $\tan\beta = 10$.  In particular, the ratio
of $0\ell$ to $1\ell$ events is still close to unity, and the $3\ell$
and $4\ell$ rates are observably large.  The main difference that
occurs at larger values of $\tan\beta$ is that there is no small
$\mu$ region.

  Since the entire mass spectrum in HENS models scales with $M_{1/2}$,
so too do the event rates.  
We have checked that for $M_{1/2}$ as large $700\;\gev$, corresponding
to a gluino mass of $m_{\tilde{g}} \simeq 1600\gev$, the inclusive
event rates in all channels other than the $0\ell$ and $4\ell$ 
are large enough that a discovery with $10 fb^{-1}$ of LHC data 
is feasible. On the other hand, the event rates become even larger 
for smaller values of $M_{1/2}$, making discovery even easier.

\section{Discussion}
\label{conc}

  Supersymmetry has been recognized as a viable theory of physics beyond
the Standard Model for many years now.  It was quickly realized that it
was not only viable, but also potentially useful in the quest to 
understand stability of the electroweak potential, radiative 
electroweak symmetry breaking, grand unification,
dark matter, and baryogenesis.  In subsequent years, there has been 
much effort devoted to understanding how supersymmetry breaking 
is to be achieved without creating additional phenomenological problems, 
such as too large FCNC.  

  There are two particularly simple alternatives 
that keep the good features of supersymmetry while (mostly) 
dismissing the bad features.  One approach is to raise the scalar masses 
significantly higher than the supersymmetric fermion masses.  
This is the idea of Split Supersymmetry discussed in the introduction.  
One drawback of this scenario is the apparent finetuning 
in the electroweak sector.  

  The approach we pursue in this article is in some sense the opposite 
of Split Supersymmetry.  Here, rather than introduce a huge hierarchy 
of scalar masses over fermion masses, we wish to zero out the 
superpartner scalar masses at some scale (i.e., ``Splat Supersymmetry").  
The simplest model of all scalar masses having a zero boundary condition 
does not work.  However, applying a small alteration to the most 
minimal idea, namely that the Higgs bosons are exempt from zero 
boundary condition requirement, preserves the good features of 
these theories, while satisfying the phenomenological requirements
described in the text.  

  This idea of Higgs Exempt No-Scale (HENS) supersymmetry has 
many phenomenological implications worthy of consideration 
at current and future experimental facilities.  For example, 
we have found that the scenario can accommodate the tantalizing
(but small) deviation of $(g\!-\!2)$ of the muon compared to the 
SM prediction.  It also suggests a near maximal leptonic signal 
for the Tevatron, and thus provides an excellent benchmark theory 
for the Tevatron to either discover this form of supersymmetry 
or rule out large regions of parameter space in a clean way.  Furthermore, 
the LHC signatures are of many multi-lepton events.  Perhaps the 
most distinctive of them is the inclusive $4\ell$ channel which 
can effectively rule out HENS models up to gaugino mass scales that 
are uncomfortably large from the normal finetuning point of view.  
For the most part, this form of supersymmetry is rather straightforward 
for the LHC to find.  An important exception is the lightest Higgs boson, 
whose mass is pressured in this scenario to be low, and thus perhaps close 
to the current bound of $114\gev$.  
Given the difficulties of finding a Higgs boson less than 
$120\gev$~\cite{Higgs Challenge}, discovering the Higgs boson might 
be one of the more challenging steps in confirming the complete 
structure of this theory.

{\bf Acknowledgements}

  We would like to thank P. Kumar for discussions, and for contributions
at the early stages of this work. We also wish to thank G. Kane,
S. Martin, A. Pierce and S. Thomas for helpful discussions.
This work is supported in part by the Department of Energy and
the Michigan Center for Theoretical Physics.

\appendix

\section*{Appendix: Sample Point Parameters}

  In this appendix, we list the relevant properties of the sample
points $A,B,C,D,E$ chosen for $\tan\beta = 10$ and $M_{1/2} = 500\;\gev$.
The locations of these points in the $SgnSqrt(m_{H_u}^2\pm m_{H_d}^2)$
plane are shown in Figs.~\ref{10-500-scan-3l} and \ref{10-500-scan-1l}.
We also list properties of the points $A',B',C',D',E'$ corresponding
to $\tan\beta = 30$ and $M_{1/2} = 500\,\gev$.

\begin{table}[hbt!]
\begin{center}
\begin{tabular}{|c|c|c|c|c|c|}
\hline
&$A$&$B$&$C$&$D$&$E$\\
\hline
&&&&&\\
$SgnSqrt(-)$ & -1480 & -1103 & -530 & -1087 & -712\\
$SgnSqrt(+)$ & -820 & -921 & -900 & -2138 & 1197\\
$\mu$ &1150 & 1033 & 868 & 1523 & 278\\
$M_{A^0}$& 1465 & 1156 & 764 & 854 & 1060 \\
$M_1$& 210&210&210&210&209\\
$M_2$& 389&389&389&389&398\\
&&&&&\\
\hline
&&&&&\\
$m_{\chi_1^0}$& 209& 209 & 209 & 210 & 193\\
$m_{\chi_2^0}$& 385 & 385 & 383 & 387 & 266\\
$m_{\chi_3^{0}}$& 1152 & 1034 & 871 & 1525 & 283\\
$m_{\chi_4^{0}}$& 1156 & 1040 & 878 & 1527 & 420\\
$m_{\chi_1^{\pm}}$& 385 & 385 & 383 & 387 & 254 \\
$m_{\chi_2^{\pm}}$& 1157 & 1041 & 878 & 1528 & 419\\
$m_{\tilde{\nu}_e}$& 223 & 274 & 315 & 281 & 304 \\
$m_{\tilde{e}_L}$& 237 & 285 & 325 & 292 & 314 \\
$m_{\tilde{e}_R}$& 384 & 312 & 223 & 308 & 249 \\
$m_{\tilde{\nu}_{\tau}}$& 217 & 272 & 315 & 288 & 298 \\
$m_{\tilde{\tau}_1}$& 221 & 261 & 214 & 261 & 233 \\
$m_{\tilde{\tau}_2}$& 383 & 328 & 331 & 352 & 310 \\
$m_{\tilde{g}}$& 1156 & 1155 & 1152 & 1161 & 1151 \\
$m_{\tilde{t}_1}$& 901 & 875 & 837 & 1027 & 719 \\
$m_{\tilde{t}_2}$& 1069 & 1046 & 1017 & 1163 & 955 \\
$m_{\tilde{u}_L}$& 1019 & 1016 & 1012 & 1007 & 1020\\
$m_{\tilde{u}_R}$& 933 & 952 & 969 & 943 & 972 \\
&&&&&\\
\hline
&&&&&\\
$\Omega h^2$& 0.098 & 0.687 & 0.096 & 0.642 & 0.134\\
&&&&&\\
\hline
\end{tabular}
\end{center}
\caption{Model parameters and particle masses for sample points
$A,B,C,D,E$, all with $\tan\beta = 10$ and $M_{1/2} = 500\,\gev$.
All dimensionful quantities in the table are listed in GeV units. 
The $\Omega h^2$ values are valid computations for the
assumption of standard thermal cosmological evolution and stable
lightest neutralino.  Viability of points $B$ and $D$ require alterations
to the standard assumptions.
\label{10-500table}}
\end{table}

\begin{table}[hbt!]
\begin{center}
\begin{tabular}{|c|c|c|c|c|c|}
\hline
&$A'$&$B'$&$C'$&$D'$&$E'$\\
\hline
&&&&&\\
$SgnSqrt(-)$ & -1595&-1019&-501&-1334&-1270\\
$SgnSqrt(+)$ & -3241&-2219&-1601&-2981&-2084\\
$\mu$ & 2195&1547&1153&2003&1542\\
$M_{A^0}$& 1031&768&522&826&1045\\
$M_1$& 212&211&211&212&211\\
$M_2$& 391&390&390&391&390\\
&&&&&\\
\hline
&&&&&\\
$m_{\chi_1^0}$& 212& 211 & 210 & 212 & 211\\
$m_{\chi_2^0}$& 390&388&387&390&388\\
$m_{\chi_3^{0}}$& 2197&1549&1156&2005&1544\\
$m_{\chi_4^{0}}$& 2197&1550&1159&2006&1546\\
$m_{\chi_1^{\pm}}$& 390&388&387&390&388\\
$m_{\chi_2^{\pm}}$& 2197&1551&1159&2006&1547\\
$m_{\tilde{\nu}_e}$& 220&287&318&256&260\\
$m_{\tilde{e}_L}$& 234&298&327&270&272\\
$m_{\tilde{e}_R}$& 403&296&219&353&342\\
$m_{\tilde{\nu}_{\tau}}$& 384&360&357&394&315\\
$m_{\tilde{\tau}_1}$& 320&272&233&324&232\\
$m_{\tilde{\tau}_2}$& 643&494&425&599&482\\
$m_{\tilde{g}}$& 1168&1162&1154&1168&1161\\
$m_{\tilde{t}_1}$& 1235&1043&924&1179&1038\\
$m_{\tilde{t}_2}$& 1393&1165&1063&1321&1160\\
$m_{\tilde{u}_L}$& 1004&1008&1008&1005&1009\\
$m_{\tilde{u}_R}$& 908&946&964&926&935\\
&&&&&\\
\hline
&&&&&\\
$\Omega h^2$& 0.105&0.521&0.0954&0.749&0.104\\
&&&&&\\
\hline
\end{tabular}
\end{center}
\caption{Model parameters and particle masses for sample points
$A',B',C',D',E'$, all with $\tan\beta = 30$ and $M_{1/2} = 500\,\gev$.
All dimensionful quantities in the table are listed in GeV units.
The $\Omega h^2$ values are valid computations for the
assumption of standard thermal cosmological evolution and stable
lightest neutralino. Viability of points $B'$ and $D'$ require alterations
to the standard assumptions.
\label{30-500table}}
\end{table}

\clearpage


\end{document}